\documentclass[twocolumn,superscriptaddress,nofootinbib]{revtex4-1}
\usepackage[utf8]{inputenc}
\usepackage{amsfonts}
\usepackage{amsmath}
\usepackage{amsthm}
\usepackage{braket}
\usepackage{graphicx}
\usepackage{hyperref}
\usepackage[dvipsnames]{xcolor}
\usepackage{amssymb}
\usepackage{dsfont}
\usepackage{verbatim}
\usepackage{amsmath,amscd}
\usepackage{pst-all}
\usepackage{leftidx}
\usepackage[all,cmtip]{xy}
	
	\usepackage{tikz,pgfplots}\pgfplotsset{compat=newest}
\usetikzlibrary{external,calc}
\newlength\figureheight
\newlength\figurewidth

\tikzset{
    double color fill/.code 2 args={
        \pgfdeclareverticalshading[%
            tikz@axis@top,tikz@axis@middle,tikz@axis@bottom%
        ]{diagonalfill}{100bp}{%
            color(0bp)=(tikz@axis@bottom);
            color(50bp)=(tikz@axis@bottom);
            color(50bp)=(tikz@axis@middle);
            color(50bp)=(tikz@axis@top);
            color(100bp)=(tikz@axis@top)
        }
        \tikzset{shade, left color=#1, right color=#2, shading=diagonalfill}
    }
}

\newcommand{\includeTikz}[2]{
\includegraphics{#1}
}

\graphicspath{{./Figures/}}

\newcommand{\qd}{\ensuremath{\mathcal{D}}}

\newcommand{\cat}{\ensuremath{\mathcal{C}}}

\newcommand{\double}[1]{\ensuremath{\mathcal{Z}({#1})}}

\newcommand{\zt}{\ensuremath{\mathbb{Z}_2}}

\newcommand{\Z}{\ensuremath{\mathbb{Z}}}
\newcommand{\C}{\ensuremath{\mathbb{C}}}

\newcommand{\layer}{\ensuremath{\ell}}
\newcommand{\dl}[1]{{\ensuremath{\bar{#1}}}}

\makeatletter
\newcommand{\vast}{\bBigg@{3}}
\newcommand{\Vast}{\bBigg@{4}}
\makeatother

\newcommand{\Ref}[1]{Ref.~\onlinecite{#1}}
\newcommand{\Refs}[1]{Refs.~\onlinecite{#1}}

\definecolor{darkblue}{RGB}{0,0,127}

\begin{document}

\psset{arrowsize=2.5pt 1.2, arrowinset=0.3}
\newpsobject{psstring}{psline}{linearc=2pt}
\newpsobject{psid}{psline}{linestyle=dotted,dotsep=2pt}
\newpsobject{pspsi}{psline}{doublecolor=lightgray, linecolor=blue, doubleline=true, linewidth=0.8pt}
\newpsobject{pssigma}{psline}{linewidth=1.5pt}

\title{
 Designer non-Abelian fractons from topological layers}
\author{Dominic J. Williamson}
\affiliation{Department of Physics, Yale University, New Haven, CT 06511-8499, USA}
\affiliation{Stanford Institute for Theoretical Physics, Stanford University, Stanford, CA 94305, USA}
\author{Meng Cheng}
\affiliation{Department of Physics, Yale University, New Haven, CT 06511-8499, USA}
\date{\today}

\begin{abstract} 
We formulate a construction of type-I fracton models based on gauging planar subsystem symmetries of topologically ordered two dimensional layers that have been stacked in three ambient spatial dimensions. 
Via our construction, any defect of an Abelian symmetry group in a two dimensional symmetry-enriched topological order can be promoted into a fracton. 
This allows us to construct fracton models supporting chiral boundaries and fractons of noninteger quantum dimension. 
We also find a lineon model supporting non-Abelian surface fractons on its boundary. 
\end{abstract}

\maketitle

\section{Introduction}
Low-energy phenomena of a stable gapped quantum phase of matter can be understood in terms of localized excitations, such as quasiparticles, and extended objects such as loops in three dimensions. Due to the energy gap, these excitations only interact topologically, manifesting in exotic exchange and braiding statistics. Recently, a new topological aspect of gapped excitations has been revealed, namely their reduced mobility even in completely translation-invariant systems, which has become the focus of intense research~\cite{nandkishore2018fractons,Pretko2020}. A rich variety of solvable models featuring excitations with reduced mobility in 3D have been discovered~\cite{Chamon2005,Haah,CastelnovoPM2012, kim20123d,YoshidaPRB2013,haah2014bifurcation, VijayPRB2015,VijayFu2017,Prem2018,Song2018,Prem2019,Bulmash2019,SSET,VijayPRB2016}.

One mechanism to produce excitations with reduced mobility is to exploit symmetries that are defined on lower-dimensional submanifolds, now known as subsystem symmetries~\cite{WilliamsonPRB2016,VijayPRB2016, you2018subsystem, ShirleyGauging2018,Williamson2018,subsystemphaserel}. It was shown that gauging planar subsystem symmetries in an Ising paramagnet yields the X-cube model~\cite{VijayPRB2016}, as well as other foliated type-I fracton models~\cite{Shirley2017,shirley2018Fractional,Slagle2018a}. Similarly, gauging a symmetry defined on a fractal submanifold in an Ising paramagnet was shown to result in type-II models such as Haah's cubic code~\cite{WilliamsonPRB2016,VijayPRB2016}. 

In this work, we explore the subsystem symmetry gauging construction further, in topologically ordered systems with \emph{higher-form} symmetries~\cite{Gaiotto2015}. For our purposes here we define a $k$-form symmetry in $d$ spatial dimensions to be a group generated by unitary operators, supported on ${(d-k)}$-dimensional (codimension-$k$) submanifolds which commute term-wise with the Hamiltonian\footnote{
We remark that our definition of higher-form symmetry differs somewhat from the common definition in the high-energy literature. In particular, we allow higher-form symmetries to be spontaneously broken, meaning they may create topologically nontrivial excitaitons at their boundaries when terminated in a finite region, and they may act nontrivially within the groundspace.
}. In this language, a conventional global symmetry is a 0-form symmetry.   Topological quantum field theories are naturally equipped with emergent higher-form symmetries.  An Abelian $(k-1)$-dimensional excitation can be created on the boundary of a $k$-dimensional operator. Such operators defined on closed submanifolds can be regarded as generators of a $(d-k)$-form symmetry. In fact, the $(d-k)$-form symmetry is spontaneously broken in the TQFT (when the energy scale is much lower than the gap to the corresponding excitation). In this work, we are primarily concerned with $1$-form symmetries in 3D. A planar subsystem symmetry can often be obtained as a special subgroup of a higher-form symmetry that is generated by elements supported on certain rigid planes.

Generally, gauging a $k$-form symmetry in a topological quantum field theory\footnote{
The $k$-form symmetry must be non-anomalous.
} results in the condensation of $(d-k-1)$-dimensional excitations that appear on the boundaries of truncated symmetry operators. This confines any excitations that have nontrivial braiding statistics with the condensate, thereby reducing the topological order to a ``smaller" one. However, in 3D if only a rigid planar subsystem symmetry subgroup of a 1-form symmetry is gauged, certain excitations in the original system can become fractonic rather than completely confined. This can be understood in terms of looplike domain-wall excitations condensing along the subsystem symmetry planes. 
This approach points us towards a construction of fractons with exotic topological properties. 

In this work, we apply this idea to explicitly construct a family of exactly solvable lattice models with non-Abelian fractons. These models are obtained by gauging subsystem symmetries in stacks of 2D non-Abelian topological phases. The 2D models exhibit 1-form symmetries, generated by Wilson loops of Abelian bosons. They can, in fact, be viewed as 2D 0-form (global) symmetry-enriched topological phases that have been gauged. The 0-form symmetry fluxes, or gauged symmetry-defects, of the 2D layers become fractons in our construction. 
In particular, we present examples with non-Abelian fractons based on layers of Ising string-net models, non-Abelian gauge theories (including $\mathbb{S}_3$, and twisted $\mathbb{Z}_2^3$, gauge theory), SU(2)$_{4k}$ theories, swap-gauged bilayer anyons, and Kitaev's honeycomb models. In the last case, the parent 1-form symmetry is actually ``anomalous", as the corresponding excitation is a fermion, but we show that the gauging construction still works.  We furthermore construct a 2D non-Abelian fracton model on the surface of a 3D abelian lineon model, by applying the gauged layer construction to the Walker-Wang model.

In section~\ref{sec:GaugedLayerConstruction} we present the single stack gauged layer construction, specialized to $\mathbb{Z}_2$ subsystem symmetries for clarity,  
in section~\ref{sec:Examples} we apply the construction to a wide range of examples including Abelian, non-Abelian and chiral layers, 
in section~\ref{sec:TDN} we describe a topological defect network construction for our models, and in section~\ref{sec:Conclusion} we compare our models to other non-Abelian fracton models in the literature. 
In appendix~\ref{sec:LWapp} we review the Levin-Wen string-net models, 
in appendix~\ref{app:notation} we introduce our notation for graded string-net models, 
in appendix~\ref{app:GaugingSubsystems} we describe the gauging procedure applied to subsystem symmetries, 
in appendix~\ref{app:GeneralModel} we present the general single stack gauged layer construction, 
and in appendix~\ref{app:honeycomblayers} we explicitly write down the Hamiltonian for the gauged honeycomb layers model. 

\section{Gauged layers construction} 
\label{sec:GaugedLayerConstruction}

In this section we first outline the general construction, before delving into required background on 1-form symmetries and finally returning to the details of the construction. 

\begin{figure}
    \centering
    \includegraphics[width=\columnwidth]{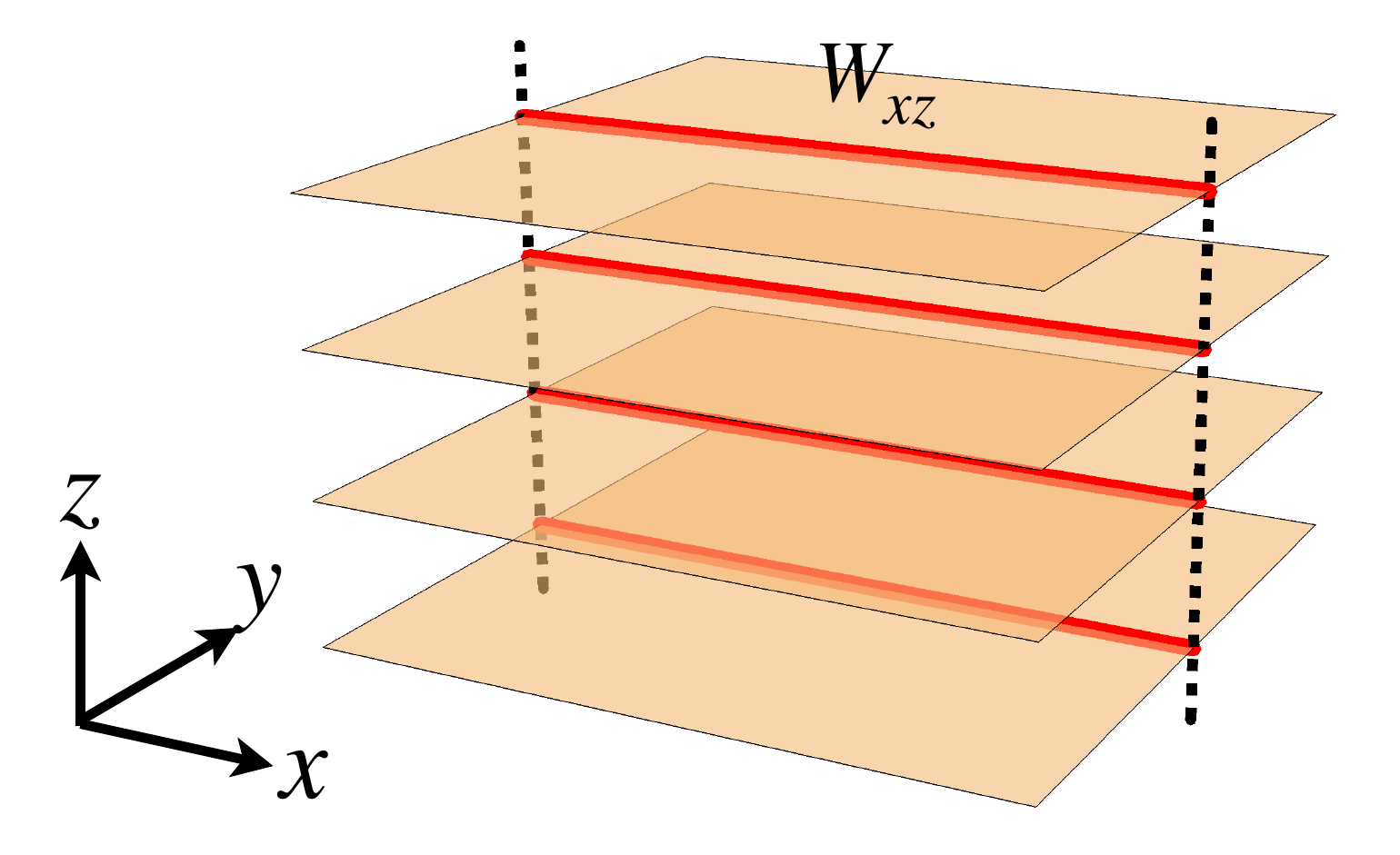}
    \caption{Illustration of a planar subsystem symmetry in the gauged layers construction. Topologically ordered 2D layers are stacked along the $\hat{z}$ direction. Each layer supports a linear subsystem symmetry, depicted in red. }
    \label{fig:layers}
\end{figure}

Consider a stack of  2D layers along the $\hat{z}$ direction. For simplicity we assume that each 2D layer lies in a gapped quantum phase of matter. Suppose that each layer respects certain 1D linear subsystem symmetries along both $\hat{x}$ and $\hat{y}$ directions. We further assume that these 1D symmetries are ``on-site", namely they can be written as tensor products of unitary operators that act on individual lattice sites. Furthermore, let us assume that these 1D symmetries form an Abelian group, for example they may correspond to the string operators of Abelian bosons or fermions. The whole 3D system has a large symmetry group given by the product of the 1D subsystem symmetry group from each layer. This large symmetry group contains a subgroup that corresponds to 2D subsystem symmetries in the $xz$ ($yz$) planes generated by the tensor product of  symmetry lines along the $\hat{x}$ ($\hat{y}$) direction in all layers. See Fig.~\ref{fig:layers} for an illustration. 

A 2D subsystem symmetry can be gauged within the defining 2D plane following the same procedure for gauging a 2D global symmetry on the lattice~\cite{Gaugingpaper,williamson2014matrix,NewSETPaper2017}. This can be carried out straightforwardly for symmetries on non-intersecting planes and when the intersecting subsystem symmetries along different directions have disjoint support. More generally when the subsystem symmetries on intersecting planes commute, they can be gauged simultaneously without any issue, see appendix~\ref{app:GaugingSubsystems}. Gauging the planar subsystem symmetry effectively couples the 2D layers together and produces a 3D phase.

\subsection{1-form symmetries in 2D topological phases}
The key building blocks in our construction are topologically ordered layers that host nontrivial anyonic excitations. In a 2D topological phase, an Abelian anyon $a$ generates a 1-form symmetry, i.e. the closed string operator $W_a$ for the anyon commutes with the Hamiltonian at low energy. Generally, these 1-form symmetries are not ``on-site"~\cite{Wen2018}. This is reflected in the possible 't Hooft anomaly obstruction to gauging the 1-form symmetries~\cite{Gaiotto2015}. The 't Hooft anomaly is nontrivial as long as the generating anyon is not bosonic, i.e. the topological twist factor $\theta_a\neq 1$.  In the condensed matter literature, gauging a 1-form symmetry is better known as anyon condensation~\cite{bais2009,kong2014}, which can only be performed for bosonic anyons (if $\theta_a=-1$, i.e. $a$ is a fermion, one can still condense $a$ by adding trivial physical fermions to the theory~\cite{aasen2017fermion}). It is believed that a non-anomalous 1-form symmetry can always be realized in a purely on-site manner. That is, the string operator is expressed as a tensor product of on-site unitaries. This is true for the fermionic case as well, where the subtle anomaly is related to the on-site operators depending on the direction of the string.  

We now discuss an alternative perspective on non-anomalous 1-form symmetries. Since the generating anyon is an Abelian boson, the underlying topological order can always be obtained by gauging a global symmetry of a symmetric phase~\cite{SET}. In fact, this symmetric phase is nothing but what remains after condensing the Abelian boson (or fully gauging the 1-form symmetry).  In general it is a symmetry-enriched topological (SET) phase, and in the special case that the condensation leaves no nontrivial deconfined anyons behind it is a symmetry-protected topological (SPT) phase. It was further shown that if the underlying topological order is non-chiral (more precisely if it has a \emph{gappable} edge), the SET phase can be realized in a symmetry-enriched string-net model~\cite{cheng2016exactly,HeinrichPRB2016,NewSETPaper2017}. Consequently, the gauged SET phase can be realized in a string-net model with an Abelian grading on the string types. 

From now on we choose a mutually commuting set of 1-form symmetry generators, corresponding to some Abelian anyons $\mathcal{A}=\{x,y,\cdots\}$. They must satisfy $M_{xy}=1$ for any $x,y\in \mathcal{A}$, where $M_{xy}$ is the braiding phase between $x$ and $y$, given by $M_{xy}=S_{xy} / S_{x1}$~\cite{Kitaev06a}. 
We denote the 1-form symmetry group thus generated also by $\mathcal{A}$. 

Given a 1-form symmetry group, excitations can be classified according to the 1-form symmetry charges of their string operators. For any anyon $a$ in the 2D gapped phases,  the charge of its string operator under a 1-form symmetry generated by $x$ is given by the mutual braiding $M_{ax}$ between $a$ and $x$.
In light of the connection to SET phases, if $M_{a x}\neq 1$, it can be viewed as a symmetry flux in the SET phase, which is an extrinsic defect.

In our construction, we only exploit a subset of the full 1-form symmetry group, i.e. its restriction to a set of rigid lines. Typically we take these lines to form a 2D grid in the plane, that effectively defines a lattice structure. We associate to the $\hat{x}$ ($\hat{y}$) direction a 1-form symmetry group $\mathcal{A}_{\hat{x}}$ ($\mathcal{A}_{\hat{y}}$). $\mathcal{A}_{\hat{x}}$ and $\mathcal{A}_{\hat{y}}$ are not necessarily the same, but we assume that their generators commute, to guarantee that both symmetries can be gauged simultaneously. 

An important distinction between gauging a ``rigid" subsystem symmetry and a full 1-form symmetry is that the anomaly-vanishing condition can be relaxed, as long as the string can be made ``on-site" it can be gauged along one direction. In particular the generating anyon may be fermionic with $\theta_a=-1$. However, being able to gauge the subsystem symmetries along two orthogonal directions requires them to commute, which means the corresponding anyons have trivial braiding statistics. 

Consider an anyon $a$ with $M_{ax}\neq 1$, i.e. the string operator $W_a$ is charged under $W_x$. To move $a$ across a line, one must apply a string operator that straddles the line which must be charged under $W_x$ applied to the line. In other words, moving the anyon $a$ across the line requires an operator that breaks the $W_x$ subsystem symmetries.

\subsection{Single stack 3D model}
\label{sec:1stack}

We now explain our construction of a 3D fracton model from a single stack of 2D topological phases that admit an abelian 1-form symmetry $\mathcal{A}$. 
Suppose the 2D topological phases are stacked along $\hat{z}$, the full system respects a large symmetry group given by the product of the 1-form symmetry $\mathcal{A}$ on each layer. This contains a subgroup corresponding to 1-form symmetries in 3D, which in turn contains a subgroup of planar subsystem symmetries parallel to $xz$ and $yz$. A fracton model is obtained by gauging these subsystem symmetries, which removes all asymmetric string operators within the layers, thus immobilizing any anyons that are moved by those operators and turning them into fractons.

To make this construction explicit, we consider 2D layers given by Levin-Wen string-net models (or suitable generalizations thereof)~\cite{qdouble,Levin2005}. In these models, degrees of freedom live on the edges of a trivalent lattice. Here we consider a lattice obtained by resolving the vertices of the square lattice to be trivalent. 
Each edge degree of freedom has a basis labeled by a finite set of string types $\{1,a,b,\dots\}$. The set $\mathcal{C}$ of string types, together with additional F-symbol data needed to consistently define the Hamiltonian, form a mathematical structure known as a unitary fusion category (UFC).  
The Levin-Wen string-net Hamiltonian 
\begin{align}
    H_\text{LW} = - \sum_{v} A_v
    - \sum_{p} B_p 
\end{align} 
consists of local commuting projector terms and is thus exactly solvable. The first type of term $A_v$, defined on vertices, enforces ``branching'' or ``fusion'' rules for the string types, i.e. only certain strings are allowed to meet at a vertex. The other type of term 
\begin{align}
    B_p =      \sum_{a\in \mathcal{C}} \frac{d_a}{\mathcal{D}^2} B_p^{a} \, ,
\end{align} 
acts on plaquettes to fluctuate the string degrees of freedom on the lattice. Here $d_a$ is the quantum dimension of an $a$ string and ${\mathcal{D}^2=\sum_a d_a^2}$ is the total quantum dimension of $\mathcal{C}$. The topological order obtained from the string-net construction is called the quantum double, or Drinfeld center, $\cal{Z}(\cal{C})$ of the UFC $\cal{C}$.
We defer a more detailed review of the string-net models to Appendix~\ref{sec:LWapp}. 

Importantly, as mentioned in the previous subsection, we assume that the string types are faithfully graded by a finite Abelian group ${\mathcal{A}}$. Namely, the set of string types $\mathcal{C}=\bigoplus_{g\in {\mathcal{A}}}\mathcal{C}_g$. For simplicity let us suppose $\mathcal{A}=\Z_2=\{0,1\}$. The generalization to other Abelian groups is straightforward and is presented in appendix~\ref{app:GeneralModel}. We associate to each edge a generalized clock operator $\widetilde{Z}_e$ that measures the grading: 
\begin{equation}
	\widetilde{Z}_e \ket{a_e}=
	\begin{cases}
		\ket{a_e} & a_e\in \mathcal{C}_0\\
		-\ket{a_e} & a_e \in \mathcal{C}_{1}
	\end{cases}.
	\label{}
\end{equation}
In particular, given the grading on string types, the branching rule must preserve the grading. In the case of a $\Z_2$ grading, it means that there can not be an allowed vertex configuration with only one string in $\mathcal{C}_{1}$. Strings in $\mathcal{C}_{1}$ must form ``loops''. This has an important consequence:  the model obeys the following 1-form symmetry
\begin{equation}
	W(\overline{\gamma})=\prod_{e\cap \overline{\gamma}}\widetilde{Z}_e \, .
	\label{eqn:Wstrnet}
\end{equation}
Here $\overline{\gamma}$ is a closed path in the dual 2D lattice. If the path is open, $W$ creates two plaquette violations on the end points, which are $\Z_2$ bosons $b$ in the topological phase. We notice that this is a special case of a general result, that is the emergent anyon theory of a string-net model $\mathcal{Z}(\mathcal{C})$ contains a subcategory of G charges (i.e. irreducible representations of $G$) when $\mathcal{C}$ is $G$-graded~\cite{ENO2009, cheng2016exactly, HeinrichPRB2016}.

To construct a 3D fracton model we consider a stack of graded string-nets, along the $\hat{z}$ direction of a cubic lattice, and gauge the 2D subsystem symmetries on dual $xy$ and $yz$ planes that are generated by products of appropriate $W$ operators on each layer as depicted in Fig.~\ref{fig:layers}. As the subsystem symmetries are defined on dual planes, we introduce $\Z_2$ gauge fields on the plaquettes of the cubic lattice. The gauge fields  for the subsystem symmetries on the dual $yz$ $(xz)$ planes are described by $X/Z_{p\hat{x}}$  $(X/Z_{p \hat{y}})$ operators, respectively. Here $X/Z$ denotes a Pauli X or Z operator. The gauge field $X/Z_{p\hat{x}}$ can be visualized as living on an edge within $p$ that is perpendicular to $\hat{x}$, and similarly for $\hat{y}$. 
Notice that if a plaquette $p$ lies in an $xy$ plane, we need to introduce both $X/Z_{p \hat{x}}$ and $X/Z_{p \hat{y}}$ as the $yz$ and $xz$ symmetries intersect there. On the other hand, plaquettes parallel to $\hat{z}$ support one gauge field each $X/Z_{p}$. More specifically, $xz$ plaquettes support $X/Z_{p \hat{x}}$ fields and $yz$ plaquettes support $X/Z_{p \hat{y}}$ fields. 

To describe the gauged model we first write down the Gauss law. For an edge $e$ along $\hat{y}$, the string can be charged under the $xz$ gauge field, so the  Gauss law is given by
\begin{equation}
A_e=\widetilde{Z}_e \prod_{p \ni e} Z_{p \hat{y}} \, .
	\label{}
\end{equation}
 For the $yz$-planar subsystem symmetry there is a similar Gauss's law for each edge along $\hat{x}$.

 We then modify the string-net Hamiltonian $H_\text{LW}$ for the 2D layers following the standard minimal coupling scheme. Within each layer, Hamiltonian terms are coupled to the gauge fields $X_{p \hat{x}}$ and $X_{p \hat{y}}$ in that plane. In fact, one only has to modify the plaquette term:
 \begin{equation}
	 B_p'= \frac{1}{\mathcal{D}^2}\Big(\sum_{a_0\in \mathcal{C}_0} d_{a_0}B_p^{a_0}+\sum_{a_{1}\in \mathcal{C}_{1}}d_{a_1}B_p^{a_1}X_{p \hat{x}}X_{p \hat{y}}\Big).
	 \label{eq:gaugedplaquetteterm}
 \end{equation}
 It is readily verified that $B'_p$ is still a projector and $[B'_p, B'_{q}]=0$ for any two plaquettes $p,q$.

Lastly, we add ``plaquette" terms for the gauge fields that enforce zero flux through each plaquette on a dual $xz$ or $yz$ plane. There are two such terms on each cube
\begin{align}
	B_c^{\hat{x}} = \prod_{p\in \partial c, p\parallel \hat{x}} X_{p \hat{x}}\,, && 
	B_c^{\hat{y}} = \prod_{p\in \partial c, p\parallel \hat{y}} X_{p \hat{y}}  \, .
	\label{}
\end{align}
It is useful to note that in the subspace where $-B_c^{\hat{x}}$ and $-B_c^{\hat{y}}$ are minimized for each cube $c$ we can write
\begin{equation}
	\prod_{p\in \partial c, p\perp \hat{z}} X_{p \hat{x}} X_{p \hat{y}}
	= \prod_{q\in\partial c, q\perp \hat{x}} X_{q \hat{x}} \prod_{q\in\partial c, q\perp \hat{y}} X_{q \hat{y}} \, .
	\label{}
\end{equation}

So the full Hamiltonian of the 3D fracton model consists of the following terms:
\begin{align}
	H&=\sum_z H_\text{LW}' -\sum_{e \perp \hat{z}} A_e-  \sum_c( B_c^{\hat{x}} + B_c^{\hat{y}})
	\nonumber \\
    &= - \sum_v A_v - \sum_{p\perp\hat{z}} B_p'  -\sum_{e \perp \hat{z}} A_e -  \sum_c (B_c^{\hat{x}} + B_c^{\hat{y}})
	\, ,
	\label{eqn:3DHamiltonian}
\end{align}
where the vertex and plaquette operators come directly from the gauged string-net Hamiltonians $\sum_zH_\text{LW}'$. 
It is straightforward to verify that all terms in the full Hamiltonian commute.

\begin{figure}
    \includegraphics[width=\columnwidth]{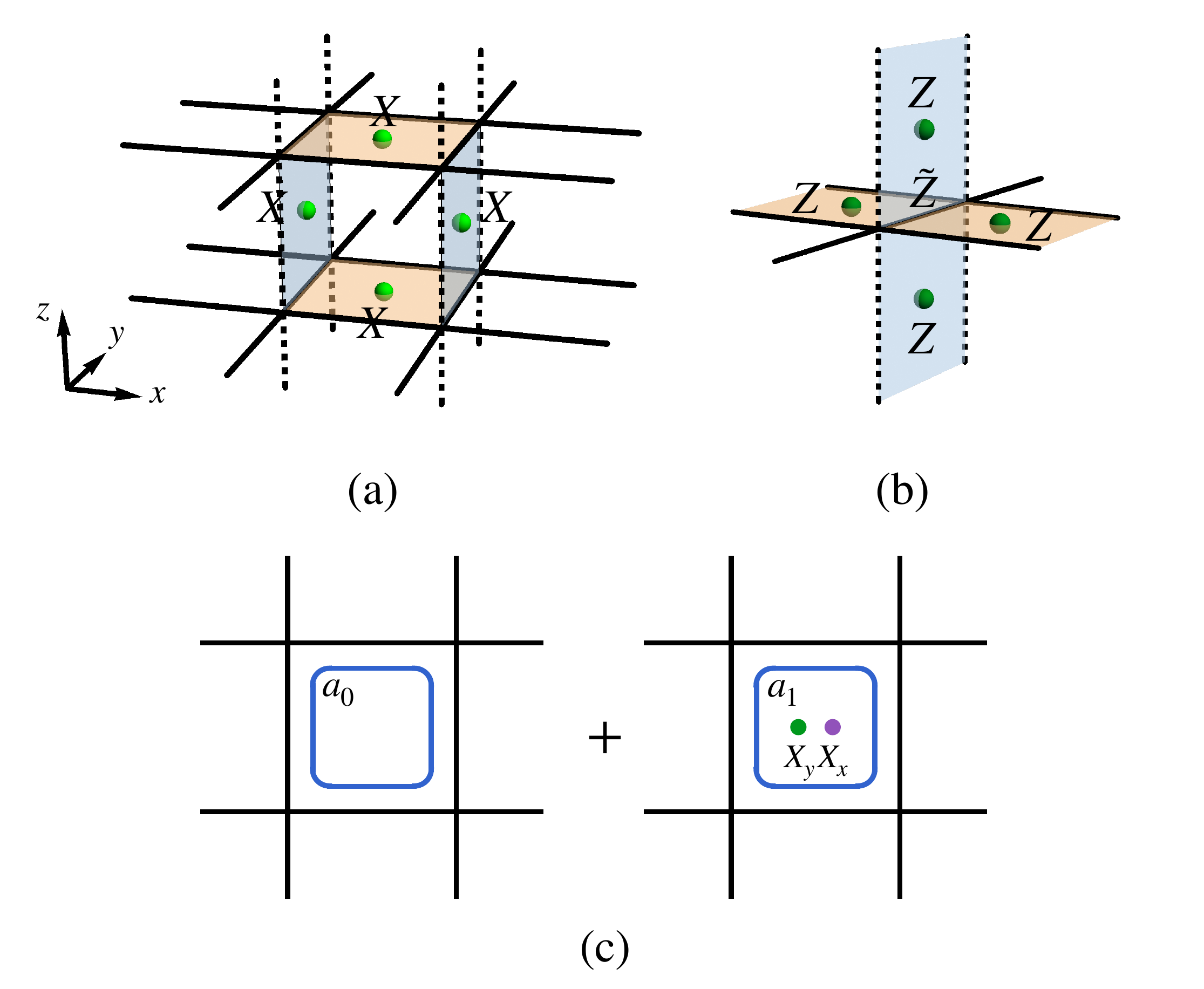}
    \caption{Illustration of $xz$ subsystem gauge fields and the Hamiltonian terms. Green dots represent the gauge fields for the $xz$ subsystem symmetry. (a) A flux constraint term. (b) The generalized Gauss's law. (c) Illustration of the gauged $B_p'$ term.
     }
    \label{fig:gauging}
\end{figure}

Allowed configurations are those that minimize the $-A_v$ and $-A_e$ terms. They consist of string-nets on lattice-edges in the 2D topological layers and $\mathbb{Z}_2$ field lines on edges formed by the intersection of the dual $xz$ and $yz$ planes  with the lattice plaquettes. The string-nets must satisfy the fusion rules of the model within each 2D topological layer. While the field lines on dual $xz$ planes must be closed, except where they end on  $\hat{x}$ edges supporting $\mathcal{C}_1$ strings, which act as sources. Similarly strings in $\mathcal{C}_1$ on edges along $\hat{y}$ act as sources for  field lines on dual $yz$ planes, which are closed otherwise. 
The ground state is given by an appropriately weighted superposition of all allowed configurations, where the weights satisfy linear relations due to the local moves induced by the $B_P'$ and $B_c$ terms. More precisely, the ground state is 
\begin{align}
    \ket{\psi_0} = \sum_{S,F} \phi(S,F) \ket{S,F} \, ,
\end{align}
where $S,F$ is an allowed configuration of strings $S$ within the topological layers, and field lines $F$ on the plaquettes. The weights are given by a product of string-net weights~\cite{Levin2005} in the layers 
\begin{align}
    \phi(S,F)=\prod_z \phi_z^{\text{LW}}(S|_z) \, ,
\end{align}
where $\phi_z^{\text{LW}}$ is the weight function for the string-net in layer $z$, and $S|_z$ is the string configuration restricted to that layer.

Next, we analyze excitations in the 3D model described by the Hamiltonian in Eq.~\eqref{eqn:3DHamiltonian}. Since Eq.~\eqref{eqn:3DHamiltonian} is a commuting projector Hamiltonian, we can build excitations from elementary violations of the local projector terms. A part of the excitation spectrum can be constructed from the excitations of the 2D layers, which are considered below. We first study excitations of the last two terms in Eq.~\eqref{eqn:3DHamiltonian}, whose properties are shared by the whole family of lattice models, regardless of the details of the layers.

Violations of cube terms $B_c^{\hat{x}/\hat{y}}$ are flux excitations of the subsystem gauge fields. Naively, one would expect these flux excitations to be mobile within their corresponding 2D planes. The string operator for a flux in the $xz$ plane is given by
\begin{equation}
	\prod_{p\in \overline{\gamma}, p \parallel \hat{y}} Z_{p \hat{y}} \, ,
	\label{}
\end{equation}
where $\overline{\gamma}$ is a path on the dual lattice in the $xz$ plane. However, generally the string operator fails to commute with $B_p'$ when the path $\overline{\gamma}$ passes through a plaquette $p$ perpendicular to $\hat{z}$. Therefore, $\overline{\gamma}$ must be a straight line along $\hat{x}$ and the corresponding excitation is an $\hat{x}$-lineon. Similarly, flux excitations of the gauge fields in the $yz$ plane are $\hat{y}$-lineons. 

We may also form a bound state of an $\hat{x}$- and a $\hat{y}$-linenon. We remark that the minimal coupling in each layer requires the subsystem gauge fields to enter the Hamiltonian via products $X_{p \hat{x}} X_{p \hat{y}}$. Hence the bound state is a $\hat{z}$-lineon, whose string operator is given by: 
\begin{equation}
    \prod_{p\in\overline{\gamma}, p \perp \hat{z}} Z_{p \hat{x}} Z_{p \hat{y}},
\end{equation}
where $\overline{\gamma}$ is a straight line along $\hat{z}$ that passes through $xy$ planes. 

We next point out that a pair of $\hat{x}$ or $\hat{y}$ lineons, adjacent to either side of an $xy$ plane, can fuse into the $\Z_2$ boson $b$ defined in Eq.~\eqref{eqn:Wstrnet} in that $xy$ plane. To see this consider multiplying adjacent $A_e$ operators, with $e \parallel \hat{y}$, along the $\hat{x}$ direction, or vice versa. Hence such a pair of lineons is an $xy$-planon.

Now we consider violations of the Gauss law term $A_e$, with $e \parallel \hat{x}$ (or $e \parallel \hat{y}$). There are two ways to create such violations: either by applying a string of $X_{p \hat{x}}$ (or $X_{p \hat{y}}$) on a dual $yz$ (or $xz$) plane, or by applying an operator to the edge that changes its grading (i.e. does not commute with $\widetilde{Z}_e$). These excitations are $yz$ (or $xz$) planons that are equivalent to a pair, or collection, of anyons adjacent to $e$ in the 2D layer.

Lastly, we turn to excitations of the 2D topological layers. Generally, the emergent topological order in a string-net model is the Drinfeld center $\mathcal{Z}(\mathcal{C})$ of the input UFC $\mathcal{C}$. Quasiparticles can be created and moved by (deformable) string operators. For a $\Z_2$-graded $\mathcal{C}$, we have written down an explicit string operator for a $\Z_2$ boson $b$. In this case, $\mathcal{Z}(\mathcal{C})$ can be viewed as a generalized $\Z_2$ gauge theory where $b$ is the gauge charge. As we have discussed earlier, quasiparticles can be classified according to their braiding with $b$, or equivalently whether their string operators are charged under the 1-form symmetry generated by $b$. 

If a quasiparticle $a$ braids trivially with $b$, the string operator for $a$ clearly commutes with all the other terms in Eq. \eqref{eqn:3DHamiltonian}, particularly the Gauss law term. Therefore $a$ remains a planon. 

If $M_{ab}=-1$, the string operator for $a$ necessarily flips the grading on each individual edge along its path, creating violations of the Gauss law term (thus creating subsystem gauge charges). In other words, this string operator is eliminated from the gauge-invariant low-energy space.  The only way to remedy this is to form a string-membrane operator, e.g. a closed string operator transporting the anyon in an $xy$ layer, attached to a membrane of $X_{p\hat{x}}X_{p\hat{y}}$ in the same layer, or a pair of such string operators in vertically separated layers, attached by a membrane of $X_p$ on the plaquettes in the $xz$ or $yz$ layers between them (notice that the two string operators involved do not have to be the same). Therefore, such anyons can only be created in quadruples. In addition, a partile-antiparticle pair of such anyons can move in the plane perpendicular to the vector connecting the two excitations. These features make them very similar to fractons in the X-cube model.

We remark that more complicated string-membrane operators can also be constructed. For example, consider three anyons $u, v$ and $w$ such that $N_{uv}^w=1$, i.e. $u\times v=w+\cdots$, and $M_{ub}=M_{vb}=-1, M_{wb}=1$. In the 2D layer before gauging, there exists an operator to split a $w$ anyon into $u$ and $v$, or a three-way junction for $u,v$ and $\overline{w}$. After gauging, the segments of the junction operator connected to $u$ and $v$ must be attached by a membrane to another string or junction operator, or they would incur an extensive energy penalty. From this perspective, two fractons separated along $\hat{z}$ form a planon, whose fusion rules and braiding statistics with other planons are inherited from the ``parent'' anyons of the 2D layers.

In summary, the gauged single stack model supports topological charges with a hierarchy of mobilities, similar to the X-cube model: 
\begin{itemize}
    \item  The anyons in the 2D layers that are charged under the 1-form symmetry generated by $b$ are promoted to fractons. 
    \item Anyons in the 2D layers that are neutral under the 1-form symmetry remain $xy$-planons. 
    \item  Gauge fluxes of the $xz$ ($yz$) planar symmetries become $\hat{x}$-lineons ($\hat{y}$-lineons), and the bound state of an $xz$ and $yz$ gauge flux becomes a $\hat{z}$-lineon. 
    \item  Gauge charges of the $xz$ ($yz$) planar symmetries become $xz$-planons ($yz$-planons), that are equivalent to a particle-antiparticle pair of fractons separated along $\hat{y}$ ($\hat{x}$). Similarly pairs of fractons separated along $\hat{z}$ become $xy$-planons.
    \item The bosonic $xy$-planon $b$ that generates the 1-form symmetry is equivalent to a pair of adjacent $\hat{x}$- or $\hat{y}$- lineons separated along $\hat{z}$. 
    Similarly a pair of $\hat{x}$- or $\hat{z}$- ($\hat{y}$- or $\hat{z}$-) lineons separated along $\hat{y}$ ($\hat{x}$) is an $xz$- ($yz$-) planon.
\end{itemize}
 See appendix~\ref{app:GeneralModel} for a derivation of these mobility constraints from emergent particle number parity conservation laws on subsystems.

A final remark: although we have worked with 2D layers described by Levin-Wen models above, the construction and our derivation of the resulting 3D fracton topological order apply to more general layers (and beyond) as long as there is an ``on-site'' 1-form symmetry, see below.

\subsection{General version of the construction}
We can carry out a similar construction with layers in two or three directions on the cubic lattice. More generally, intersecting layers whose 1-form symmetries combine to give planar subsystem symmetries. 
The layered structure suggests such models can be defined given a foliation~\cite{Shirley2017}, although a foliation structure is not necessary. 

In fact, we can carry out the construction whenever there is a planar subsystem symmetry. In particular any 1-form symmetry in 3D will contain such a subgroup. Hence the construction also applies to 3D gauge theory with Abelian flux loops and graded Walker-Wang models~\cite{williamson2016hamiltonian}. 
In particular, it was shown in Ref.~\onlinecite{Williamson2018a} that gauging planar subsystem symmetries of the 3D toric code results in the X-cube model. We have similarly verified that gauging planar subsystem symmetries of the 3D toric code with fermionic charge (the Walker-Wang model~\cite{walker2012} based on sVec) also leads to the X-cube model, up to local unitary equivalence.

\section{Examples} 
\label{sec:Examples}

In this section we present a series of examples demonstrating the versatility of the gauged layer construction. We first consider layers of Abelian gauge theory, allowing for 3-cocycle twists, including toric code and double semion models. We find models that are not foliated equivalent to X-cube but share the same gauge structure. 
Next, we consider several non-Abelian examples, including layers of Ising and Tambara-Yamagami string-nets, $\mathbb{S}_3$ and non-Abelian twisted $\mathbb{Z}_2^3$ gauge theory, SU(2)$_{4k}$ and swap-gauged bilayer theories. 
Finally we consider the example of Ising anyons in Kitaev's honeycomb model, where the 1-form symmetry is anomalous, and on the surface of a Walker-Wang model. 
Interestingly, these examples lead to fracton models with gapless chiral boundaries, exotic non-Abelian fractons that do not have square root integer quantum dimensions, and non-Abelian fractons on the 2D surface of a lineon model.

\subsection{Abelian models}

First we present several Abelian examples.

\subsubsection{$\Z_2$ toric code} 
In this section we carry out the gauged layers construction for layers of 2D $\Z_2$ toric code~\cite{qdouble} along the $\hat{z}$ direction, with 1-form symmetry generated by $m$ (or equivalently $e$). We find a model that is equivalent to the X-cube model. 

Consider a 2D toric code Hamiltonian on the square lattice, where each edge has a qubit:
\begin{equation}
	H_{\text{2DTC}} = - \sum_{v} \prod_{e \ni v} \widetilde{Z}_e 
	- \sum_{p} \prod_{e \in \partial p}  \widetilde{X}_e \, ,
	\label{}
\end{equation}
with a 1-form symmetry given by $\widetilde{Z}_e$'s acting along closed loops in the dual lattice. As usual, we refer to the vertex violations as $e$ and plaquette violations as $m$. The 1-form symmetry is generated by closed $m$ strings.

Following our procedure we stack the 2D toric codes along the $\hat{z}$ direction and gauge the $\zt$ planar subsystem symmetry subgroup within the product of the 1-form symmetry groups from each layer. This results in a model 
\begin{widetext}
\begin{align}
    H = 
    - \sum_{z} 
    \left(
	\sum_{v\in \ell_z} \prod_{e \ni v, e\perp \hat{z}} \widetilde{Z}_e
	+ \sum_{p \in \ell_z}   X_{p\hat{x}} X_{p\hat{y}} \prod_{e \in \partial p}  \widetilde{X}_e
	+ \sum_{e \in \layer_z} \widetilde{Z}_e \prod_{p \ni e, p \perp \hat{z} } Z_{p{\hat{e}}} \prod_{p \ni e, p \not\perp \hat{z} } Z_{p}
    \right)
    \nonumber \\
    -\sum_{c} \left( \prod_{p \in \partial c,\, p \perp \hat{z}} X_{p\hat{x}} \prod_{p \in \partial c,\, p \perp \hat{y}} X_{p}
    +
    \prod_{p \in \partial c,\, p \perp \hat{z}} X_{p\hat{y}} \prod_{p \in \partial c,\, p \perp \hat{x}} X_{p}
    \right)
    \, .
\end{align}
Here $z$ is the layer index and $\ell_z$ denotes the $z$-th layer. We have also used a slightly modified convention for subsystem gauge fields: for plaquettes $p$ perpendicular to the layers, we suppress the $\hat{x}/\hat{y}$ index as it is uniquely determined by the plaquette. For $xy$ plaquettes we make the subsystem index explicit. 

From now on we write $\widetilde{X}_e$ as $X_e$ for notational clarity.  
To simplify the model, we apply a circuit consisting of controlled-$X$ gates given by
\begin{align}
    U = \prod_{p \perp \hat{z}} \prod_{ e \in \partial p} CX_{p{\hat{e}},e}
    \,
    \prod_{p \not\perp \hat{z}} \prod_{ e \in \partial p,\, e \perp \hat{z}} CX_{p,e}
    \, .
\end{align}
The resulting Hamiltonian is given by
\begin{align}
   U H U^\dagger= 
    - \sum_{z} \left(
    \sum_{v\in \layer_z} \prod_{e \ni v} Z_e
    \prod_{p \in \layer_z,\, v\in\partial p} Z_{p\hat{x}} Z_{p\hat{y}}
    \prod_{p \not\in \layer_z,\, v\in\partial p} Z_p
    + \sum_{p\in \layer_z}   X_{p\hat{x}} X_{p\hat{y}} 
    + \sum_{e \in \layer_z} Z_{e} 
    \right)
    \nonumber \\
    -\sum_{c} \left( \prod_{p \in \partial c,\, p \perp \hat{z}} X_{p \hat{x}} \prod_{p \in \partial c,\, p \perp \hat{y}} X_{p}
    +
    \prod_{p \in \partial c,\, p \perp \hat{z}} X_{p\hat{y}} \prod_{p \in \partial c,\, p \perp \hat{x}} X_{p}
    \right) 
    \, .
\end{align}

Now we consider the subspace where $Z_e=1$  and $ X_{p_x} X_{p_y} = 1$. Physically, this means the generalized Gauss law is strictly enforced, and there are no plaquette excitations in each layer (so the 1-form symmetry is topological).
We define the remaining qubit on plaquettes in the $xy$-plane by $ X_{p \hat{x}} \sim X_{p \hat{y}} \mapsto Z_p $ and $Z_{p \hat{x}} Z_{p \hat{y}}\mapsto X_p$.  The Hamiltonian in this subspace is phase equivalent to the unrestricted model and is given by 
\begin{align}
     U H U^\dagger \mapsto
    - 
    \sum_{v} \prod_{p \ni v} X_p
    -\sum_{c} \left( 
    \prod_{p \in \partial c,\, p \not\perp x} Z_{p}
     +
     \prod_{p \in \partial c,\, p \not\perp y} Z_{p}
    \right)
    \, ,
\end{align}
which is in fact the X-cube Hamiltonian on the dual lattice. 
\end{widetext}

We have also applied the gauged layer construction to $\mathbb{Z}_2$ planar subsystem symmetries of 2D toric code layers stacked along all three axial directions, forming a cubic lattice with two spins per edge. There we found a gauged model that is equivalent to two copies of the X-cube model. 

\subsubsection{Other Abelian examples}
We now briefly discuss gauged layer constructions of several more complicated Abelian fracton phases. 
\newline

\noindent \textbf{$\mathbb{Z}_4$ and twisted $\mathbb{Z}_2\times\mathbb{Z}_2$ gauge theories:}
The example in the preceding section can be generalized straightforwardly to $\mathbb{Z}_N$ toric code layers. Furthermore, when $N$ contains a prime raised to a power greater than 1 there may be inequivalent choices of anomaly-free 1-form symmetries that lead to different models via the gauged layer construction. In particular, for $\mathbb{Z}_4$ toric code layers one may instead gauge $\mathbb{Z}_2\times\mathbb{Z}_2$ planar subsystem symmetries generated by $e^2$ and $m^2$ string operators, where $e$ and $m$ are the generators of the $\Z_4$ electric and magnetic charges, respectively.

To find a simple lattice realization we take the layers to be twisted $\mathbb{Z}_2\times\mathbb{Z}_2$ gauge theories with string types labelled by group elements $A=a_0a_1$, where $a_0,a_1\in \mathbb{Z}/2\mathbb{Z}=\{0,1\}$, and fusion rules (group multiplication) are denoted additively. The $F$-symbols are given by the so-called type-II 3-cocycle~\cite{Propitius1995}
\begin{align}
   F^{ABC}_{A+B+C}= \alpha(A,B,C)=(-1)^{a_0b_1c_1}\, .
\end{align}
This UFC is denoted as $\mathrm{Vec}_{\Z_2\times\Z_2}^\alpha$.
We make use of the well known fact~\cite{Propitius1995} that $\mathbb{Z}_2\times\mathbb{Z}_2$ gauge theory, twisted by a type-II 3-cocycle is equivalent to $\mathbb{Z}_4$ gauge theory, or more precisely $\mathcal{Z}(\text{Vec}_{\mathbb{Z}_2\times\mathbb{Z}_2}^{\alpha})\cong \mathcal{Z}(\text{Vec}_{\mathbb{Z}_4})$. 
The obvious $\mathbb{Z}_2\times \mathbb{Z}_2$ grading on the string types corresponds to the 1-form symmetry group generated by $e^2$ and $m^2$ in the emergent $\mathbb{Z}_4$ guage theory~\cite{NewSETPaper2017}. 
The gauged layer construction then results in Abelian fractons generated by $e$ and $m$. In particular fermionic and semionic anyons such as $em^2$ and $em$, respectively, are promoted to fractons. 
This model is not foliated equivalent to $\mathbb{Z}_2 \times \mathbb{Z}_2$ X-cube as the fusion of two stacks of $e$ fractons along $z$ yields a pair of irreducible lineons at its end points. This follows form the fact that fusing two $e$ fractons results in an $e^2$ planon, which is equivalent to a pair of irreducible lineons. 

We remark that this is a different twisted generalization of the X-cube model than that in Ref.~\onlinecite{Shirley2019d}, although both are based upon the 1-form symmetry, or Lagrangian algebra object, generated by $e^2$ and $m^2$ in $\mathbb{Z}_4$ toric code. 

The anomaly free $\mathbb{Z}_2\times\mathbb{Z}_2$ 1-form symmetry in this example can also be used to construct an anisotropic gauged layer model where $\mathbb{Z}_2$ subsystem symmetries on $xz$ planes generated by $e^2$ string operators and on $yz$ planes generated by $m^2$ string operators are gauged. This results in an anisotropic lineon model where $e$ is promoted to an $\hat{x}$-lineon and $m$ is promoted to a $\hat{y}$-lineon. We remark that this is distinct from the anisotropic lineon models in Ref.~\onlinecite{Fuji2019}. 
\newline

\noindent \textbf{Doubled semion model:} 
Similar to the twisted layers in the above example, the double semion is a twisted $\mathbb{Z}_2$ gauge theory with $\mathbb{Z}_2$ string types and $F$-symbols given by the so-called type-I 3-cocycle 
\begin{align}
   F^{abc}_{a+b+c}=(-1)^{abc}\, .
\end{align}
The emergent topological order in each layer decouples into a chiral and anti-chiral semion theory, each of which supports a single nontrivial anyon type corresponding to a semion $s$, with $\theta_s=i$, and an anti-semion $\overline{s}$, with $\theta_{\bar{s}}=-i$, respectively. 
The $\mathbb{Z}_2$ grading on the string types corresponds to the 1-form symmetry group generated by $s\overline{s}$ in the layers. Gauging the planar subsystem symmetries of these layers results in a fracton model with similar fusion structure to X-cube, but where the semion $s$ is promoted to a fracton. 
This is distinct from the previous semionic generalization of the X-cube model in Ref.~\onlinecite{MaPRB2017}, which was shown to be foliated equivalent to the standard X-cube model~\cite{shirley2018Fractional}. 
It would be interesting to evaluate whether our model is foliated equivalent to X-cube or not.
\newline 

\noindent \textbf{Twisted $\Z_N$ gauge theory:} 
The previous example generalizes straightforwardly to layers of twisted $\Z_N$ gauge theory~\cite{dijkgraaf1990, Propitius1995}. There are $N$ different choices for the 3-cocycle twist, labeled by an integer $p=0,1,\dots, N-1,$ as $\mathcal{H}^3[\Z_N, \mathrm{U}(1)]=\Z_N$.  
These twisted gauge theories can be realized in generalized Levin-Wen models~\cite{LinPRB2014}. The input UFC is denoted $\mathrm{Vec}_{\Z_N}^{\alpha}$. This category has $N$ string types $a=0,1,\cdots, N-1$, with $\Z_N$ fusion rules $a\times b=[a+b]$, and $F$-symbols given by the 3-cocycle~\cite{Moore89b, Propitius1995}:
\begin{equation}
	F^{abc}_{a+b+c}=\alpha(a,b,c)=e^{\frac{2\pi p}{N^2}a(b+c-[b+c])}\, ,
	\label{}
\end{equation}
where square brackets denote mod $ N$. 

The emergent anyons are labeled by tuples $(a, m)$ of gauge charge $a$ and gauge flux $m$. Their fusion rules are 
\begin{equation}
	(a,m)\times (b, n)=\Big([a+b+\frac{2p}{N}(m+n-[m+n])], \big[m+n\big]\Big). 
	\label{}
\end{equation}
In particular 
\begin{equation}
    (0,1)^{\times N}=([2p],0).
\end{equation}
Physically, this means each unit gauge flux carries a fractional $\Z_N$ charge when $[2p]\neq 0$. 
Hence the anyons have the following topological twist factors 
\begin{equation}
	\theta_{(a,m)}=e^{\frac{2\pi i}{N}am}e^{i\frac{2\pi p}{N^2}m^2}.
	\label{}
\end{equation}
We remark, for odd $N$ and $(p,N)=1$, the fusion rules imply that the fusion ring of the anyon theory is $\Z_{N^2}$.

The input UFC admits an obvious $\Z_N$ grading, and the corresponding $\Z_N$ boson is the gauge charge $(1,0)$. 
Applying the gauged layer construction yields an Abelian fracton model where all the $(0,m)$ fluxes become fractons. 
This model can be viewed as a twisted variant of the X-cube model, distinct from other twisted generalizations that have previously appeared in the literature~\cite{MaPRB2017,Song2018,YYZ2018,Shirley2019d,Devakul2020}, to the best of our knowledge. 
For $[2p]\neq 0$, this gauged layer model is not foliated equivalent to the $\mathbb{Z}_N$ X-cube model. To demonstrate this, we point out that fusing $N$ fractons results in a $([2p],0)$ gauge charge, which is equivalent a pair of $x$ or $y$ lineons, denoted $\ell$. Formally we write 
\begin{equation}
	(0,1)_{j}^{\times N}=\ell_{j-\frac{1}{2}}^{2p} \ell_{j+\frac{1}{2}}^{N-2p} .
	\label{}
\end{equation}
Here $j$ is the layer index.
Hence fusing $N$ stacks of such fractons results in a pair of isolated irreducible lineons
\begin{equation}
\prod_{j=j_{d}}^{j_u}	(0,1)_{j}^{\times N}= \ell_{j_d-\frac{1}{2}}^{2p} \ell_{j_u+\frac{1}{2}}^{N-2p} ,
	\label{}
\end{equation}
which is not possible in a model that is foliated equivalent to the $\mathbb{Z}_N$ X-cube model.

\subsection{Non-Abelian fractons}
\label{sec:ising}
Here we consider several non-Abelian UFCs which can be directly fed into the gauged layers construction to produce models with non-Abelian fractons. 

\subsubsection{Doubled Ising model}

A simple non-Abelian UFC is the Ising category, with three types of strings, $1, \sigma$ and $\psi$. Their fusion rules are given by
\begin{equation}
	\begin{gathered}
	\sigma\times\sigma=1+\psi,\\
	\sigma\times\psi=\sigma,\\
	\psi\times\psi=1.
	\end{gathered}
	\label{}
\end{equation}
The $\mathbb{Z}_2$ grading is given by $\mathcal{C}_0 = \{1,\psi\},~\mathcal{C}_{1}=\{\sigma\}$.  The 2D string-net construction produces a doubled Ising topological phase Ising$\times\overline{\text{Ising}}$, with nine quasiparticles $a\overline{b}$ where $a,b\in\{1,\sigma,\psi\}$ ($1$ is suppressed below). This is because the input category admits modular braidings. In particular, the $\psi\overline{\psi}$ particle is a $\mathbb{Z}_2$ boson, associated with the $\mathbb{Z}_2$ grading of the Ising category. 
The remaining seven nontrivial anyons can be divided into two groups: $\psi, \overline{\psi}$,  $\sigma\overline{\sigma}$ are neutral under the 1-form symmetry $W_{\psi\overline{\psi}}$ so they remain planons, while $\sigma,\overline{\sigma}, \sigma\overline{\psi}, \psi\overline{\sigma}$ are charged, and so become fractons in the 3D gauged model. The general discussion above implies that a ``dipole'' of $\sigma$'s separated along $\hat{z}$ is a planon that can move freely in the $xy$ plane.

 \begin{figure}[t]
	 \centering
	 \includegraphics[width=\columnwidth]{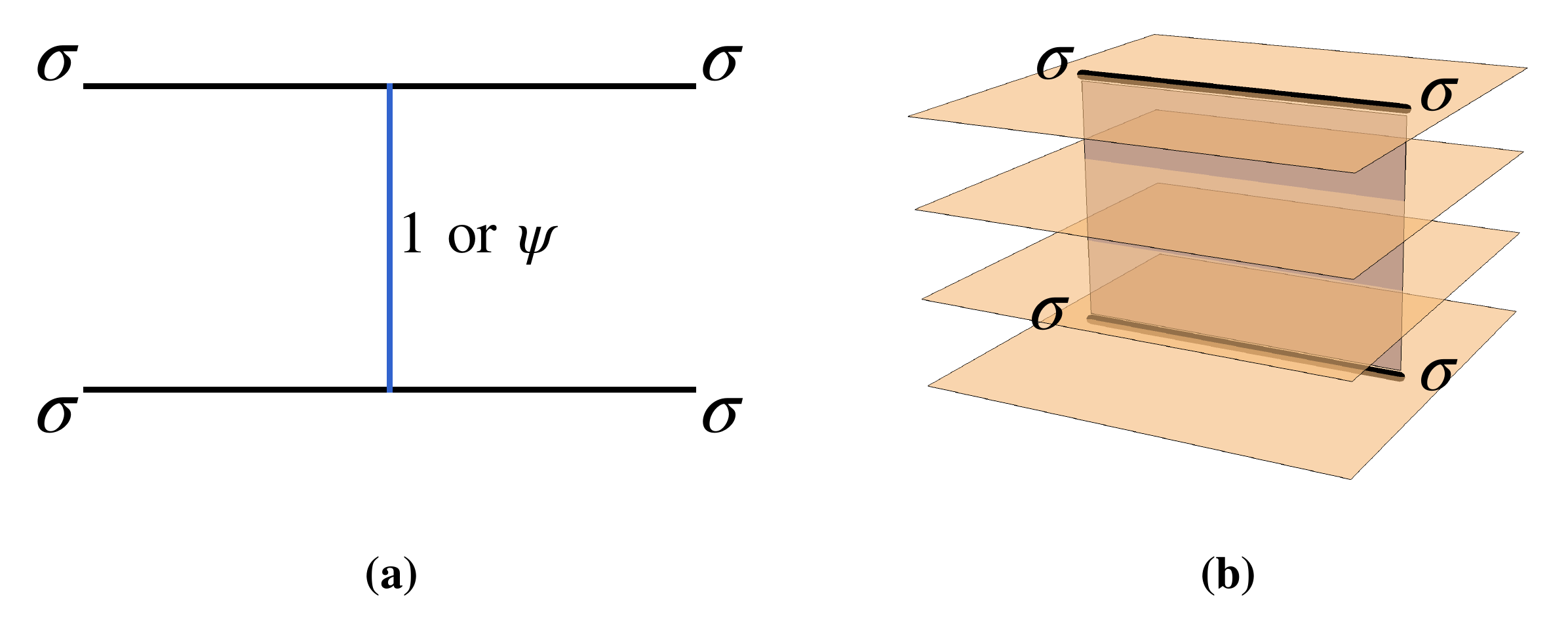}
	 \caption{String and membrane operators that created non-Abelian Ising $\sigma$ particles. 
	 (a) String operators that create the two orthogonal basis states in the fusion space of four $\sigma$ anyons. 
	 (b) A string-membrane operator that creates four $\sigma$ fractons in the Ising gauged layer model. 
	 }
	 \label{fig:ising}
 \end{figure}

 Now let us clarify what it means to have ``non-Abelian'' fractons. In two dimensions, an anyon $a$ has a quantum dimension $d_a\geq 1$. If $d_a>1$, the anyon is non-Abelian.  When $n$ identical, well-separated anyons $a$ are created on a sphere (such that the total topological charge is trivial), the dimension of the degenerate excited state subspace grows asymptotically as $d_a^n$ as $n\rightarrow \infty$. The different excited states are locally indistinguishable, and can be labeled by the corresponding fusion trees. As an example, in the doubled Ising model the ground state dimension for $n$ Ising anyons $\sigma$ is $2^{n/2-1}$, and therefore $d_\sigma=\sqrt{2}$. In the minimal case $n=4$, the string operators that create the two states are shown schematically in Fig.~\ref{fig:ising}(a). A pair of $\sigma$'s either fuse into the vaccum or $\psi$ and with four $\sigma$'s there is a choice of the intermediate fusion channel being $1$ or $\psi$. When the fusion channel is $1$, one simply applies two string operators of $\sigma$ to create the four anyons. When the fusion channel is $\psi$, the more complicated operator as depicted in Fig. \ref{fig:ising}(a) is needed, where a $\psi$ string connects the two $\sigma$ strings.

 Now we come back to the 3D gauged model. It can easily be seen that both operators, after attaching a membrane in the $xy$ plane, can still create the same configurations of $\sigma$'s. Therefore the two-fold topological degeneracy of four Ising anyons in the same $xy$ plane remains intact. We can also make sense of non-Abelian braiding: suppose the four $\sigma$ are in the $j$-th layer. Even though a single $\sigma$ is immoble, as we have mentioned in Sec. \ref{sec:1stack} a $\sigma$-dipole $\sigma_j\sigma_{j+1}$ can move in the $xy$ plane. Braiding such a dipole around the other anyons induces non-Abelian transformation in the two-dimensional space. The exact form of the transformations can be directly read off from that of the non-Abelian braiding of Ising anyons in 2D.
 
 More generally, if $n$ Ising anyons are in the same $xy$ plane then there is still a $2^{n/2-1}$ fold topological degeneracy. On the other hand, if the membrane is perpendicular to the $xy$ plane, the four anyons as shown in Fig. \ref{fig:ising}(b) have a unique state associated with them. Therefore the degeneracy for the non-Abelian fractons depends on the precise configuration of them. The same is true for the other fractons $\overline{\sigma}, \sigma\overline{\psi}, \psi\overline{\sigma}$.

We can also show that what we construct is an intrinsically 3D fracton phase. In particular, it can not be transformed to a simpler, ``Abelian'' fracton phase (e.g. X-cube model) stacked with decoupled 2D layers of non-Abelian topological phases. If that were the case, there must be at least one irreducible Abelian fracton excitation, which is absent in our model. 

\subsubsection{Other non-Abelian models}

Here we briefly discuss a few more examples of non-Abelian fracton phases obtained via the gauged layer construction. 
\newline

\noindent \textbf{$\Z_N$ parafermions:} 
A generalization of the Ising category is the so-called $\Z_N$ Tambara-Yamagani category where $N$ is an odd integer. There are $N+1$ labels, the first $N$ of which are basically elements of $\Z_N$, denoted by $[j],~j=0,1,\dots, N-1$. The last one $\sigma$ satisfies
\begin{equation}
	\sigma\times\sigma=[0]+[1]+\cdots+[N-1].
	\label{}
\end{equation}
The $\Z_2$ grading is given by $\mathcal{C}_0=\{[0],[1],\dots, [N-1]\}$ and $\mathcal{C}_1=\{\sigma\}$. For a certain choice of the $F$ symbol, the Drinfeld center has the same topological order as the theory Spin$(N)_2\times\overline{\mathrm{SU}(N)}_1$~\cite{SET, NewSETPaper2017}. We thus find that gauging $\Z_N$ Tambara-Yamagani string-net layers yields ``parafermionic'' fractons with quantum dimension $\sqrt{N}$. 
\newline

\noindent\textbf{$\mathbb{S}_3$ gauge theory:} 
With this example we point out that the gauged layer construction allows one to simply produce a non-Abelian fracton model from layers of $\mathbb{S}_3$ gauge theory, see Ref.~\onlinecite{beigi2011quantum} for a review. Here the group is generated by $\{s,r|s^2=r^3=1, srs=r^{-1}\}$.
In the construction we gauge the subsystem symmetry generated by copies of the nontrivial $\zt$ boson corresponding to the sign representation of the $\mathbb{S}_3$ group. This forces the two inequivalent fluxes labelled by the conjugcay class of size three to become non-Abelian fractons of quantum dimension $3$. 
For the lattice model we take strings labelled by elements of $\mathbb{S}_3$, with $\mathbb{Z}_2$ grading given by $\mathcal{C}_0=\{ 1,r,r^2\} $, $\mathcal{C}_1=\{ s,sr,sr^2 \}$, where the fusion rules are given by group multiplication and the $F$-symbols are trivial. The same construction can be easily generalized to all dihedral groups.
\newline

\noindent
\textbf{Non-Abelian twisted $\mathbb{Z}_2^3$ gauge theory:} 
A  twisted  $\mathbb{Z}_2^3$ checkerboard model with non-Abelian fractons was previously presented in Ref.~\onlinecite{Song2018}, here we explain an analogous construction via gauged layers. 
The layers are taken to be twisted $\mathbb{Z}_2^3$ gauge theories with string types labelled by group elements $A=a_0a_1a_2$, where $a_0,a_1,a_2\in\mathbb{Z}_2$, and $F$-symbols given by the nontrivial so-called type-III 3-cocycle 
\begin{align}
   F^{ABC}_{A+B+C}= \alpha(A,B,C)=(-1)^{a_0b_1c_2}\, .
\end{align}
It is well known that the emergent topological order in the 2D layers is equivalent to $\mathbb{D}_4$ gauge theory, or more technically $\mathcal{Z}(\text{Vec}_{\mathbb{Z}_2^3}^{\alpha})\cong \mathcal{Z}(\text{Vec}_{\mathbb{D}_4})$, 
see Ref.~\onlinecite{Propitius1995}. 
The string types trivially admit a $\mathbb{Z}_2^3$ grading, which corresponds to the $\mathbb{Z}_2^3$ group of abelian bosons in the emergent $\mathbb{D}_4$ gauge theory. Gauging the subsystem symmetries generated by these bosons results in the remaining fourteen inequivalent non-Abelian anyons in each layer, which all have quantum dimension $2$, being promoted to fractons. 
\newline

\noindent \textbf{SU(2)$_{4k}$ anyons:} 
Another family of examples are given by the SU(2)$_{4k}$ categories. The $4k+1$ anyon types are labeled by an SU(2) spin $j=0, 1/2, \cdots, 2k$. In particular,  the $j=1/2$ anyon is always non-Abelian, with quantum dimension $d_{1/2}=2\cos \frac{\pi}{4k+2}$, and the highest-spin anyon $j=2k$ is a $\Z_2$ boson, which braids nontrivially with all anyons of half-odd-integer spin. The complete set of algebraic data can be found in \Ref{Bonderson2007}.  Applying the gauged layer construction to the planar symmetries generated by the $2k$ boson, all anyons of half-odd-integer SU(2) spin anyons are promoted to fractons. It is worth pointing out that in this way one obtains non-Abelian fractons whose squared quantum dimensions are not integers for all $k>1$. For example, for $k=2$, it is known that the SU(2)$_8$ theory is actually equivalent to bilayer Fibonacci anyons where the $\Z_2$ layer exchange symmetry has been gauged~\cite{SET}. In that case the spin-$1/2$ anyon, which has quantum dimension $\sqrt{\frac{5+\sqrt{5}}{2}}$, is promoted to a fracton. 

We remark that SU(2)$_{4k}$ anyon theories have chiral edge states, with chiral central charge $c_-=\frac{6k}{2k+1}$. 
The gauging does not change the chiral central charge, so the 3D model also has chiral edge modes on side surfaces. 
Technically the bulk anyon theory only determines the chiral central charge of the edge theory modulo 8. However, it turns out that the edge of a stack with an arbitrary number of SU(2)$_{4k}$  theories with $k>4$ cannot be gapped\footnote{More precisely, the Witt class of SU(2)$_k$ is infinite-order as long as $k\neq 1, 2, 4$}~\cite{Davydov2013a}. 
Hence any model supporting the bulk fracton theory given by gauged SU(2)$_{4k}$ layers must have a gapless edge for $k>4$. 

To construct an explicit commuting projector lattice model that supports a fracton whose squared quantum dimension is not an integer, one could instead apply the gauged layers construction to SU(2)$_k$ string-net Hamiltonians. This results in a similar, though nonchiral, model to the one described in this section. 
\newline

\noindent
\textbf{Swap-gauged bilayer anyons:} 
For any bilayer topological phase of the form $\cal{B}\times \cal{B}$ where $\cal{B}$ is an anyon theory, there is a $\Z_2$ symmetry that swaps the two layers. A $\Z_2$ defect of the symmetry $X_0$ satisfies
\begin{equation}
    X_0\times X_0=\sum_{a\in \cal{B}}(a,a).
\end{equation}
Therefore the quantum dimension of $X_0$ is equal to the total quantum dimension $\cal{D}$ of $\cal{B}$. Other defects can be obtained from $X_0$ by attaching anyons in either layer: $X_a=X_0\times (a,1)$.

After gauging the $\Z_2$ swap symmetry, the defects become symmetry fluxes $X_a^\pm$ where $\pm$ denotes $\Z_2$ charge. We can then apply the gauged layer construction to the planar symmetries generated by $\mathbb{Z}_2$ gauge charges in the swap-gauged bilayer theory.
This promotes $X_a^\pm$ to fractons.  For example, the Fibonacci anyon theory has only one nontrivial topological excitation $\tau$, in addition to the vacuum $1$. They satisfy fusion rules
\begin{align}
    \tau \times \tau = 1 + \tau \, ,
\end{align}
which implies the quantum dimension of $\tau$ is $d_\tau=\phi$, and the total quantum dimension is $\mathcal{D}^2=\phi \sqrt{5} $, where $\phi=(1+\sqrt{5})/2$ is the golden ratio. 
Hence, as mentioned above, applying the gauged layer construction to the swap-gauged bilayer Fibonacci theory produces a fracton with quantum dimension $\sqrt{\frac{5+\sqrt{5}}{2}}$. 
This may have interesting implications for braiding universality using fractons~\cite{Bruillard2017}.

\subsection{Fermionic symmetries in the honeycomb model}

We now discuss a concrete lattice example where the subsytem symmetry being gauged corresponds to  string operators that create fermions. As we will see, as far as straight line operators are concerned, they still take the form of a tensor product of an on-site unitary operator, and can be gauged following the same procedure.

 The model we consider in each layer is Kitaev's honeycomb lattice model~\cite{Kitaev06a}:
\begin{equation}
	H=-\sum_{\alpha=x,y,z}J_\alpha \sum_{\alpha\text{-links}}\sigma_i^\alpha \sigma_j^\alpha .
	\label{eq:honeycomb_model}
\end{equation}
It can be mapped to fermions coupled to static $\mathbb{Z}_2$ gauge fields. This is achieved by the following Majorana representation of a spin-$1/2$:
\begin{align}
	\sigma^\alpha=ib^\alpha c\, , && \alpha=x,y,z,
	\label{}
\end{align}
with the gauge constraint $b^xb^yb^z c=1$. The Hamiltonian then becomes
\begin{equation}
	H=\frac{i}{2}\sum_{\langle ij\rangle}J_{\alpha_{ij}}u_{ij}c_ic_j.
	\label{eqn:kitaev_H_2}
\end{equation}
Here $u_{ij}=ib^{\alpha_{ij}}_ib^{\alpha_{ij}}_j$ represent static $\mathbb{Z}_2$ gauge fields. 
The phase diagram is worked out in Ref. \onlinecite{Kitaev06a}. When any one of the couplings dominates, e.g. $J_x \gg J_y,J_z$, the model is in a gapped toric code phsae. 
For isotropic couplings $J_x=J_y=J_z$, the fermions are gapless while the $\Z_2$ fluxes remain gapped (this is called the $B$-phase). If a suitable time-reversal breaking Zeeman field is then turned on, a mass gap opens for the fermions, taking the model into a non-Abelian topological phase with Ising topological order~\cite{Kitaev06a}.

\begin{figure}
    \centering
    \includegraphics[width=\columnwidth]{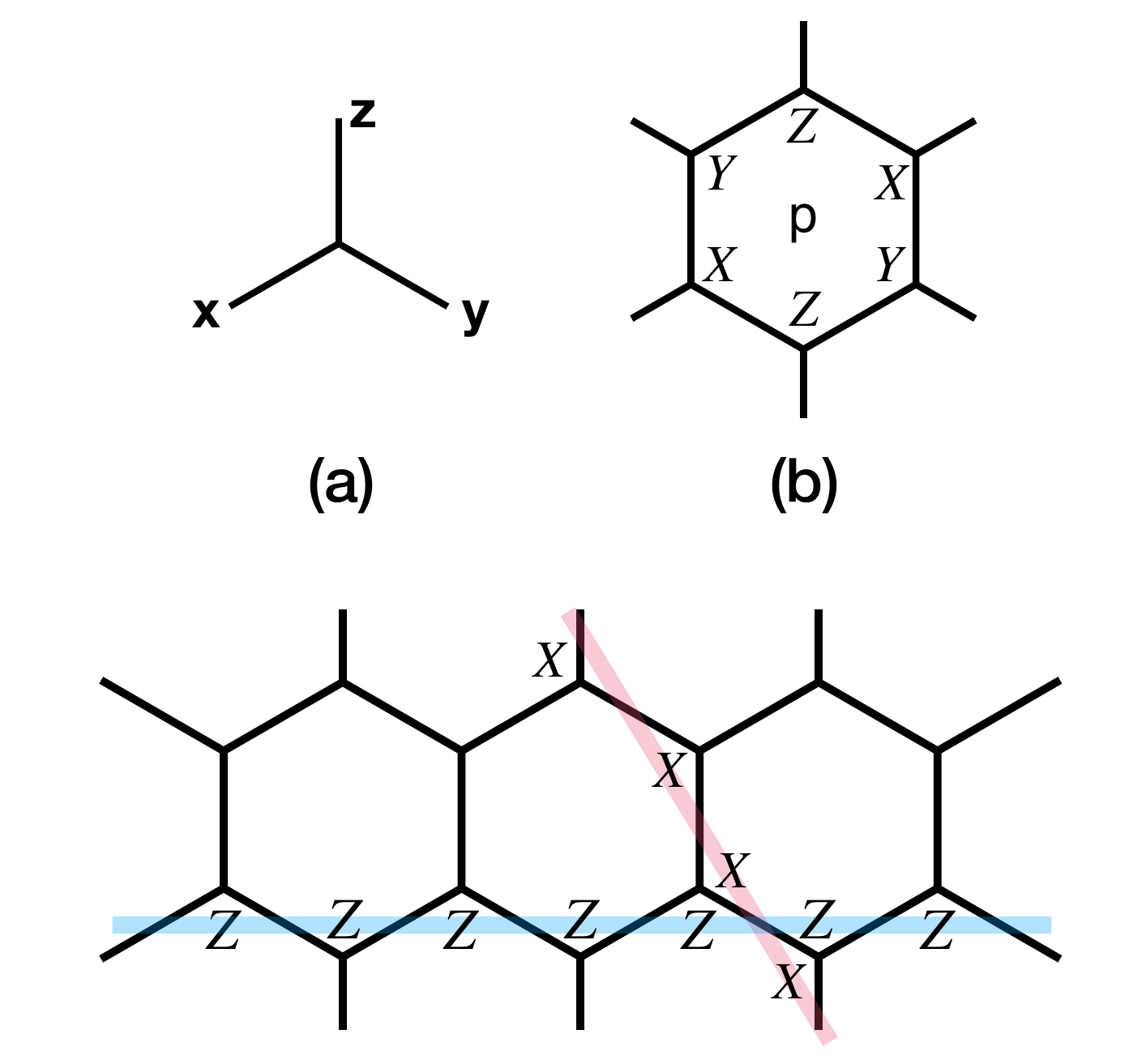}
    \caption{Fermionic subsystem symmetries in Kitaev's honeycomb lattice model.}
    \label{fig:honeycomb}
\end{figure}

The $\Z_2$ gauge fluxes are measured by Wilson loop operators. Following the notation in \Ref{Kitaev06a}, define
\begin{equation}
	K_{ij}=\sigma_i^\alpha \sigma_j^\alpha, \: \text{if }ij\text{ is a $\alpha$-link}.
	\label{}
\end{equation}
Then for a path $\gamma$ connecting sites $j_0, j_1, \cdots, j_n$, 
\begin{equation}
	W_f(\gamma)=K_{j_nj_{n-1}}\cdots K_{j_1j_0}=\prod_{s=1}^n (-iu_{j_s j_{s-1}})c_nc_0.
	\label{}
\end{equation}
Thus $W_f(\gamma)$ creates a fermion at each end point of $\gamma$.
For a closed path $\gamma$, $j_n\equiv j_0$, it can be shown that $W_f(\gamma)$ commutes with the Hamiltonian throughout the whole phase diagram, and thus generates a 1-form symmetry.
The minimal symmetry generators are $W_p$ on each plaquette, measuring $\Z_2$ flux in the hexagon. In the ground state all $W_p=1$.  $W_f(\gamma)$ restricted to a straight line $\gamma$ generates a 1D subsystem symmetry of the model. In Fig.~\ref{fig:honeycomb} two such subsystem symmetries are shown.

The gapped $B_\nu$ phase can be reached while preserving the fermionic 1-form symmetry via the perturbation
\begin{align}
    V= \sum_{\langle ij \rangle\langle ik \rangle} K_{ij} K_{ik} + \sum_{\langle ij \rangle\langle ik \rangle\langle i\ell \rangle} K_{ij}K_{ik}K_{i \ell}  ,
\end{align} 
where $\langle ij \rangle$ denotes a pair of adjacent vertices and $i,j,k,\ell$ are all distinct. Perturbing the honeycomb Hamiltonian in Eq.~\eqref{eq:honeycomb_model}, to $H-\Delta V$, leads to the gapped Ising phase with chirality ${\nu=\text{sgn} \Delta}$. 

By utilizing the $X^{\otimes L}$ and $Z^{\otimes L}$ linear subsystem symmetries of the 2D perturbed Kitaev honeycomb model, shown in Fig.~\ref{fig:honeycomb}(c), we can apply the gauged layers construction to obtain a 3D Hamiltonian. Details of the lattice Hamiltonian are presented in Appendix~\ref{app:honeycomblayers}. 

Most interestingly, if the layers are in the gapped non-Abelian chiral Ising  $B_\nu$ phase, the 3D model supports Ising fractons. This is somewhat similar to the model described in Sec.~\ref{sec:ising}, except the gauged honeycom layers model also exhibits chiral Majorana surface modes. 
Alternatively, the layers can be tuned to the gapless $B$-phase, in which case the gapped $\Z_2$ fluxes remain fractons, but the (planon) fermions become gapless. 
Finally, for layers tuned to one of the toric code phases, we expect a model equivalent to the X-cube model. 
This presents a family of models exhibiting nontrivial phase transitions, chirality and non-Abelian fractons, which deserve further exploration. We remark that these features differentiate our construction from the previous models built from honeycomb layers in Refs.~\onlinecite{PhysRevB.96.165106,Fuji2019}.

\subsection{Surface fractons from the Ising Walker-Wang model} 

The Walker-Wang model is a generalization of the Levin-Wen model to 3D, based upon braided fusion categories~\cite{walker2012} (which are algebraic theories of anyons). For a modular tensor category (MTC) $\mathcal{M}$ the topological order in the bulk is trivial, while there is a canonical gapped boundary to vacuum that supports potentially chiral $\mathcal{M}$ anyons, despite the fact that the Hamiltonian is a sum of local commuting projector terms. 
A (necessarily Abelian) grading on the string types of a Walker-Wang model induces a 1-form symmetry analogously to our discussion of graded Levin-Wen models. 
Such 1-form symmetries act via products of on-site operators, that are diagonal in the string label basis, and hence can always be gauged~\cite{WW1Form}. 
For MTCs there is a universal Abelian grading induced by braiding with the Abelian anyons in the MTC. In this case, where the 1-form symmetry corresponding to an abelian anyon $a$ ends on a boundary it implements the string operator $W_a$. 
This can be used to effectively implement even anomalous 1-form symmetries on the boundary. 

Consider a semi-infinite Walker-Wang model based on the Ising MTC, $\mathbb{Z}_2$-graded by the Abelian anyon group generated by $\psi$, with the canonical smooth boundary at $z=0$. We gauge the subsystem symmetries generated by $\psi$ on $xz$ and $yz$ planes, see section~\ref{sec:ising}.  In the bulk this results in a lineon topological order equivalent to the twice foliated X-cube model~\cite{shirley2018Fractional}. 
On the boundary surface there is a nontrivial 2D fracton model, where the Ising $\sigma$ anyons are promoted to fractons. 
Similar to the gauged layer construction, a pair of $\sigma$ anyons on the surface separated by $\hat{x}$ ($\hat{y}$) is an $xz$ ($yz$) planon gauge charge. Gauge fluxes of the $xz$ ($yz$) planar symmetries are $xz$ ($yz$) planons, as expected in this case. However, the naive string operator for these planons must be dressed by a $\psi$ string, making them fermions. At the boundary surface all the gauge fluxes are equivalent to the $\psi$ anyon, which remains a planon there. 
This surface is quite remarkable as it has recently been argued that there are no purely 2D gapped fracton topological orders~\cite{Aasen2020}, and it is distinct to the surfaces found in Ref.~\onlinecite{Bulmash2018b} as there are no fractons in the bulk in this case. It remains to be seen whether it is possible to have a nontrivial 2D fracton order on the surface of an invertible 3D bulk.

This model has a straightforward construction via a topological defect network, this also holds for our more general models (see below). 
The model where only the $xz$ planar symmetries are gauged is equivalent to coupling layers of 2D toric code on the $xz$ planes to the Ising surface theory via domain walls that are formed, after folding the Ising surface, by condensing the $\psi \overline{\psi}$ boson in Ising$\times\overline{\text{Ising}}$. 
Similarly the full gauged layer model is equivalent to coupling layers of 2D toric code on $xz$ and $yz$ planes, with a certain choice of gapped boundary where they intersect, to an Ising anyon theory surface in a similar fashion. 
Hence the same fracton surface order could be achieved by gauging a layer of the honeycomb model, in the gapped Ising phase, stacked on the surface of trivial symmetric bulk degrees of freedom, although the Hamiltonian would not be commuting projector with this alternate construction.

\section{Topological defect network interpretation}
\label{sec:TDN}

Recently it has been proposed that any fracton model admits a description by a network of topological defects within a conventional topological order~\cite{Aasen2020} (see also Refs.~\onlinecite{Wen2020} and~\onlinecite{JWang2020} for related constructions). This includes 2D layers with topological order, coupled along gapped boundaries, within a trivial ambient 3D topological order as a special case~\cite{Wen2020}. Here we demonstrate that the wide range of models obtained via the gauged layers construction are indeed described by topological defect networks, supporting the conjecture in Ref.~\onlinecite{Aasen2020}. 

The topological defect network construction of a gauged layer model is obtained by introducing a 2D Abelian $A$ gauge theory onto each $A$ subsystem symmetry plane of the ungauged layers, along with appropriate gapped boundary conditions where these Abelian gauge theories intersect the layers. Where different planes of Abelian gauge theory intersect they simply pass through one another.

Let us describe the nontrivial gapped boundary where the $A$ gauge theory intersects a topological layer. Recall that by assumption, the anyon theory for the layer $\mathcal{C}$ contains a set of bosons that form an Abelian group $A$. Therefore $\cal{C}$ admits an $\widehat{A}$ grading induced by braiding phases with $A$-bosons
\begin{align}
\mathcal{C}=\bigoplus_{\chi \in \widehat{A}} \mathcal{C}_\chi ,
\end{align}
where $\widehat{A}$ denotes the character group. 
The gapped boundary is specified by the following rules: anyons in $\mathcal{C}_0$ can pass through, while anyons in $\mathcal{C}_\chi$ can only pass through by creating a $\chi$ gauge charge in the intersecting $A$ gauge theory. Similarly gauge charges in the $A$ gauge theory are allowed to pass through the boundary, while $g$ gauge fluxes can only pass through by creating a $g$ $A\text{-boson}$ in $\mathcal{C}$. 
The gapped corner terms, where two planes of $A$ gauge theory and a $\mathcal{C}$ layer all intersect, are specified by the gauged string-net plaquette terms which couple to both intersecting gauge fields, see Eq.~\eqref{eq:gaugedplaquetteterm}.

\begin{figure}
    \centering
    \includegraphics[width=\columnwidth]{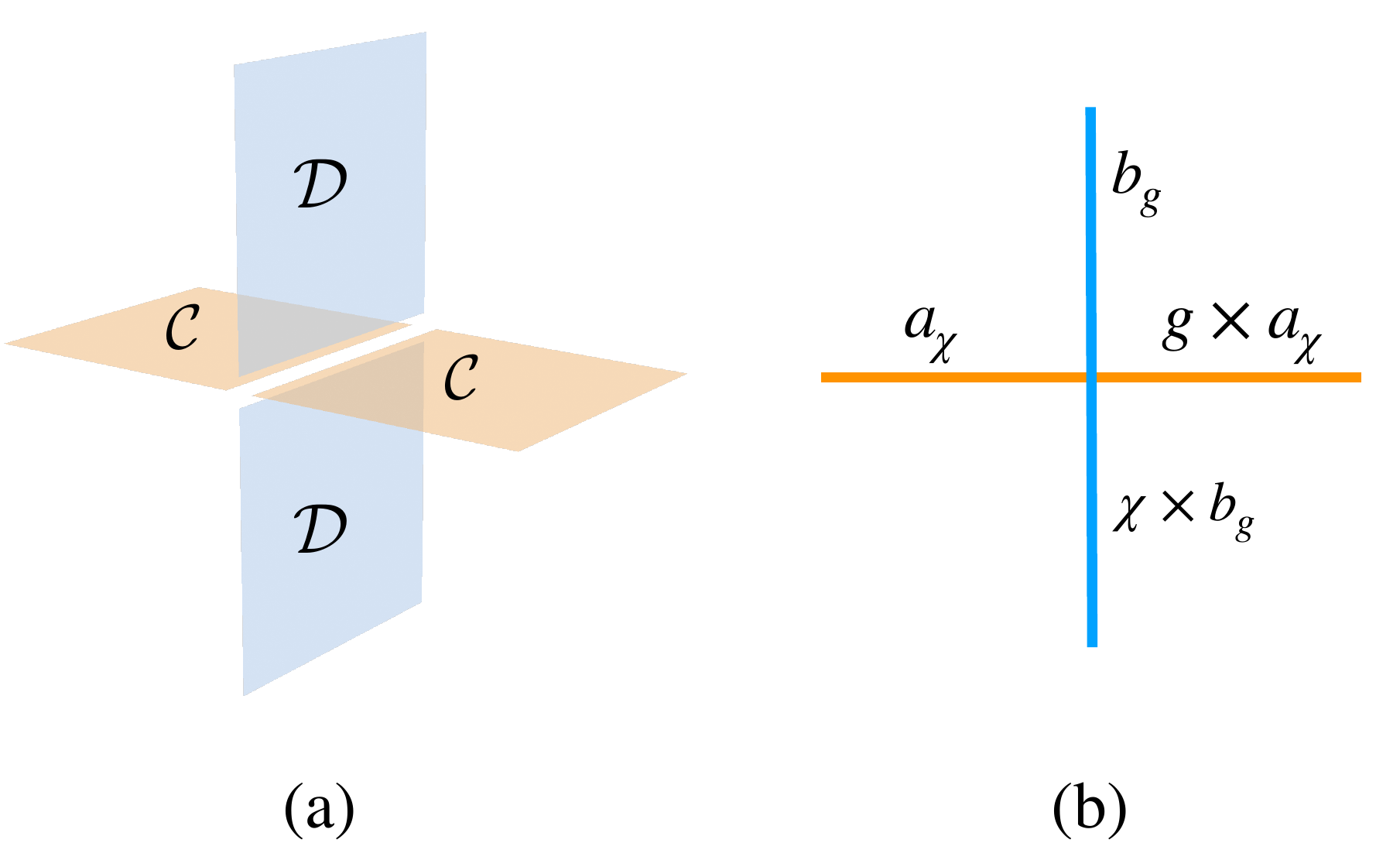}
    \caption{Illustration of the defect network construction. (a) The intersection of $\cal{C}$ and $\cal{D}$ layers, which can be cut into four semi-infinite layers and folded as the boundary of the 2D topological phase $\cal{C}\boxtimes \overline{\cal{C}}\boxtimes\cal{D}\boxtimes\overline{\cal{D}}$. (b) The Lagrangian algebra for the gapped boundary.}
    \label{fig:dn}
\end{figure}

We remark that the gapped boundary where an $A$ gauge theory $\mathcal{D}=\mathcal{Z}(\text{Vec}_A)$ intersects a $\mathcal{C}$ layer is equivalent to a gapped boundary to vacuum of $\mathcal{C}\boxtimes \overline{\mathcal{C}}\boxtimes\mathcal{D}\boxtimes\overline{\mathcal{D}}$ via folding  where $\overline{\mathcal{C}}$ denotes orientation-reversal of $\cal{C}$, as illustrated in Fig. \ref{fig:dn}(a). Hence the gapped boundary is specified by a Lagrangian algebra object~\cite{Levin2013,kong2014} in $\mathcal{C}\boxtimes \overline{\mathcal{C}}\boxtimes\mathcal{D}\boxtimes\overline{\mathcal{D}}$. 
Before presenting the appropriate object, we first point out that $\mathcal{D}$ trivially admits a grading into flux sectors
\begin{align}
    \mathcal{D} = \bigoplus_{g\in A} \mathcal{D}_g \, ,
\end{align}
which can also be though of as induced by braiding phases with the $\widehat{A}$ gauge charges. 
The Lagrangian algebra object for a trivial intersection of $\mathcal{C}$ and $\mathcal{D}$ layers is given by 
\begin{equation}
    \mathcal{L}=\sum_{a\in \mathcal{C}}\sum_{b\in \mathcal{D}}(a,\overline{a},b,\overline{b}) \, ,
\end{equation}
while for the nontrivial gapped boundary in the gauged layers topological defect network it is given by 
\begin{align}
\label{eq:LAO}
    \mathcal{L}=\sum_{a\in\mathcal{C}}\sum_{b\in\mathcal{D}} \sum_{\chi \in \widehat{A}} \sum_{g \in A} 
    (a_\chi , \overline{g} \otimes \overline{a}_{\overline{\chi}}, \chi \otimes b_g, \overline{b}_{\overline{g}})  \, .
\end{align}
Fig. \ref{fig:dn}(b) shows the anyon condensation in the unfolded intersection.

This point of view reveals a further generalization of the gauged layer construction by replacing the $A$ gauge theory with an arbitrary $A$-graded anyon theory $\mathcal{D}$. In fact, the Lagrangian algebra object in Eq.~\eqref{eq:LAO} remains the same for this general case. The lattice model can also be generalized to this case by replacing the $A$ spins on dual $xz$ and $yz$ planes with $A$-graded string-net models. 
Moreover, any model obtained by gauging an Abelian planar subsystem symmetry can equivalently be obtained by introducing Abelian gauge theory layers coupled to the original system via a topological defect network. All such models then admit a further generalization where the gauge theory layers are replaced by graded string-net models.

\section{Discussion and conclusion}
\label{sec:Conclusion}

In this work we introduced a new class of type-I fracton models, constructed by gauging planar subsystem symmetries inherited from Abelian 1-form symmetries of layered 2D topological orders. 
Our models are capable of hosting vastly more general types of non-Abelian fractons as well as chiral boundaries, as demonstrated through examples including gauged layers of Ising string-net, $\mathbb{S}_3$ and twisted $\mathbb{Z}_2^3$ gauge theory, SU(2)$_{4k}$ anyons, swap-gauged bilayer anyons and Kitaev's honeycomb model. We also demonstrated that applying our construction to the Ising Walker-Wang model leads to non-Abelian surface fractons on the boundary of a 3D planon model. 

Our models highlight the potential existence of fracton models with even more exotic behaviours than those discovered to date. In particular, the existence of non-Abelian fractons with noninteger squared quantum dimension, which may have important applications for topological quantum computation. 
A notable interesting open question is whether such non-Abelian fractons can be realized in a type-II fracton model, and  whether a model of this type can serve as a self-correcting quantum memory. 
This could be addressed by searching for fractal subsystem symmetries of some convention non-Abelian topological order, or more generally within the topological defect network formalism~\cite{Aasen2020}. 

We close by comparing and contrasting our construction with other non-Abelian fracton models in literature.

In \Ref{VijayFu2017} Vijay and Fu constructed a non-Abelian fracton model by coupling $p_x+ip_y$ superconducting layers to the Majorana checkerboard model. This is similar in spirit to our construction (in particular the gauged honeycomb layers model in the gapped $B_\nu$ phase), in the sense that a $\mathbb{Z}_2$ symmetry flux (i.e. vortices in the $p_x+ip_y$ superconductors) is promoted into a fracton by coupling layers to subsystem gauge fields. However, the construction in \Ref{VijayFu2017} relies heavily on the specific geometry of the Majorana checkboard model. Our construction can be regarded as a certain generalization of this model.

In \Ref{Song2018}, Song et al. took a different approach by ``twisting'' spin checkerboard models by 3-cocycles. The resulting model for $\mathbb{Z}_2^3$ with a type-III cocycle twisting, on $xy$ planes only, supports non-Abelian fractons with integer quantum dimension $2$. The non-Abelian fractons in our twisted $\mathbb{Z}_2^3$ example are very similar to those in \Ref{Song2018}, it would be interesting to establish a local unitary equivalence between the models, possibly up to stacking with X-cube and 2D toric code layers. Our models are more general in the sense that they allow for layers beyond twisted gauge theory, and non-Abelian fractons with noninteger quantum dimension. 

Another general construction of 3D fracton models from intersecting 2D layers is through the mechanism of ``p-loop condensation"~\cite{Vijay2017,MaPRB2017}. \Ref{Prem2018} applied this construction to layers of doubled SU(2)$_k$ theories, generating 3D models that feature Abelian fractons and non-Abelian lineons, which is fundamentally different from our construction. It is an interesting open question whether a certain generalized string-membrane-net model~\cite{Slagle2018a}, which combines the two constructions, allows for the most general foliated type-I fracton models with non-Abelian excitations. 

Our construction is analogous to independent p-loop condensations within the gauged planes rather than the whole 3D bulk. To demonstrate this we point out that gauging a symmetry condenses its domain walls. Hence p-loop condensation is equivalent to gauging a 1-form symmetry inherited from the 1-form symmetries in each layer that are generated by the string operators for the particles attached to the p-loops. Our construction instead gauges a planar subsystem symmetry inherited from the linear subsystem symmetry subgroup of the 1-form symmetry in each layer
\begin{align}
    &G_{\text{planar subsystem}} \leq G_{\text{3D 1-form}} \leq G_{\text{stacked 2D 1-forms}} \, ,
    \nonumber
    \\
    & \hspace{.7cm} \texttt{"} \hspace{.5cm} \leq G_{\text{stacked 2D linear subsystem}}  \leq \hspace{.5cm} \texttt{"} \hspace{.7cm}
    \, .
\end{align}

Recently, another route leading to non-Abelian fractons was presented in \Refs{Bulmash2019, Prem2019}. One starts from multiple copies of a fracton model, such as the X-cube model, and gauges a global layer-permutation symmetry. The resulting models exhibit ``panoptic" topological order, including non-Abelian fractons that always have integer quantum dimensions. In addition, there are completely mobile particles (gauge charges) and loop excitations, which were absent in the constructions discussed above. We expect that the swap-gauged bilayer X-cube model can be obtained by gauging a $\mathbb{Z}_2\times\mathbb{Z}_2$ planar subsystem symmetry of the 3D toric code, see \Ref{SSET} for a proposed construction of this kind. 
It remains an open problem to generalize our gauged layer construction to capture panoptic models such as the swap-gauged bilayer cubic code from \Refs{Bulmash2019, Prem2019}, which contain non-Abelian fractons that are created at the corners of fractal operators. 

Even more recently, a model with Non-Abelian fractons of quantum dimension $2$ was constructed in \Ref{Aasen2020} using a topological defect network based on 3D $\mathbb{D}_4$ gauge theory. Like the gauged bilayer fracton models, this model also supports mobile abelian gauge charges and non-Abelian loop excitations. This model can be obtained by gauging subsystem symmetries of a 3D $\mathbb{D}_4$ gauge theory, which is a special case of the generalized string-membrane-net models that will be present in a future work.

\section{Acknowledgements}

D. W. and M.C. would like to thank X. Chen for discussions and explaining unpublished works. D. W. acknowledges useful discussion with D. Aasen, A. Dua and A. Prem. 
M. C. is supported by NSF CAREER (DMR-1846109) and  the Alfred P. Sloan
foundation. D. W. acknowledges support from the Simons Foundation.

\bibliographystyle{apsrev4-1}
\bibliography{refs}

\clearpage 

\appendix 

\section{Levin-Wen models}
\label{sec:LWapp}

The string-net model takes as input a unitary fusion category, which describes a finite set of string types that fuse in a non-associative way, captured by the so-called $F$-symbols~\cite{Levin2005}. 
The Hamiltonian is a sum of two types of terms:
\begin{equation}
    H=-\sum_v A_v - \sum_p B_p.
\end{equation}
For clarity of presentation, we consider the Hamiltonian on a honeycomb lattice with undirected edges and no fusion multiplicity, although these restrictions can be dropped via a suitable generalization. 
The vertex terms $A_v$ ensure that fusion rules are obeyed at each vertex:
\begin{equation}
	A_v\vast|
	\begin{array}{c}
	\includeTikz{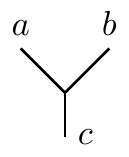}{
	\begin{tikzpicture}[baseline={($ (current bounding box) - (0,0pt) $)}, scale=0.9]
		\draw[thick] (0, .5) node [above] {$a$}--(.5, 0) ;
		\draw[thick] (1, .5) node [above] {$b$}--(.5, 0) ;
		\draw[thick] (.5, 0)--(.5, -.5) node [right] {$c$};
	\end{tikzpicture}
	}
	\end{array}
	\vast\rangle
	=N_{ab}^c\vast|
	\begin{array}{c}
	\includeTikz{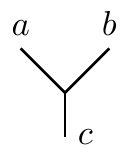}{
	\begin{tikzpicture}[baseline={($ (current bounding box) - (0,0pt) $)}, scale=0.9]
		\draw[thick] (0, .5) node [above] {$a$}--(.5, 0) ;
		\draw[thick] (1, .5) node [above] {$b$}--(.5, 0) ;
		\draw[thick] (.5, 0)--(.5, -.5) node [right] {$c$} ;
	\end{tikzpicture}
	}
	\end{array}
	\vast\rangle.
	\label{}
\end{equation}
Here $N_{ab}^c=1$ when $c$ appears as a fusion channel of $a$ and $b$, otherwise $0$. The plaquette terms take the following form:
\begin{equation}
    B_p=\frac{1}{\mathcal{D}^2}\sum_s d_s B_p^s \, ,
\end{equation}
where $d_s$ is the quantum dimension of string type $s$, and $\mathcal{D}$ is the total quantum dimension. 
The $B_p^{s}$ operators have the following graphical representation: imagine adding a loop of type $s$ onto the plaquette, and fusing it into the edges 
\begin{equation}
	B_p^s
	\vast|\,
	\begin{array}{c}
		\includeTikz{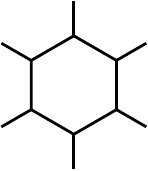}{
	 \begin{tikzpicture}[baseline={($ (current bounding box) - (0,1.5pt) $)}]
		 \draw[thick] ({sqrt(3)/4},0.25) -- (0,0.5) -- ({-sqrt(3)/4},0.25) -- ({-sqrt(3)/4},-0.25) -- (0,-0.5) -- ({sqrt(3)/4},-0.25) -- ({sqrt(3)/4},0.25);
	\draw[thick] ({sqrt(3)/4},0.25) -- ({1.7*sqrt(3)/4},1.7*0.25);
	\draw[thick] (0,0.5) -- (0,1.7*0.5);
	\draw[thick] ({-sqrt(3)/4},0.25) -- ({-1.7*sqrt(3)/4},1.7*0.25);
	\draw[thick] ({-sqrt(3)/4},-0.25) -- ({-1.7*sqrt(3)/4},-1.7*0.25);
	\draw[thick] (0,-0.5) -- (0,-1.7*0.5);
	\draw[thick] ({sqrt(3)/4},-0.25) -- ({1.7*sqrt(3)/4},-1.7*0.25);
\end{tikzpicture}
} 
	\end{array} \,
	\vast\rangle
	=
	\vast|\,
	\begin{array}{c}
	\includeTikz{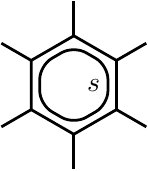}{
	 \begin{tikzpicture}[baseline={($ (current bounding box) - (0,1.5pt) $)}]
		 \draw[thick] ({sqrt(3)/4},0.25) -- (0,0.5) -- ({-sqrt(3)/4},0.25) -- ({-sqrt(3)/4},-0.25) -- (0,-0.5) -- ({sqrt(3)/4},-0.25) -- ({sqrt(3)/4},0.25);
\draw [rounded corners, thick] ({0.8*sqrt(3)/4},0.8*0.25) -- (0,0.8*0.5) -- ({-0.8*sqrt(3)/4},0.8*0.25) -- ({-0.8*sqrt(3)/4},-0.8*0.25) -- (0,-0.8*0.5) -- ({0.8*sqrt(3)/4},-0.8*0.25) -- cycle;
	\draw[thick] ({sqrt(3)/4},0.25) -- ({1.7*sqrt(3)/4},1.7*0.25);
	\draw[thick] (0,0.5) -- (0,1.7*0.5);
	\draw[thick] ({-sqrt(3)/4},0.25) -- ({-1.7*sqrt(3)/4},1.7*0.25);
	\draw[thick] ({-sqrt(3)/4},-0.25) -- ({-1.7*sqrt(3)/4},-1.7*0.25);
	\draw[thick] (0,-0.5) -- (0,-1.7*0.5);
	\draw[thick] ({sqrt(3)/4},-0.25) -- ({1.7*sqrt(3)/4},-1.7*0.25);
	\node  at (0.2, 0) {\scalebox{0.8}{$s$}};
\end{tikzpicture}
} 
	\end{array} \,
	\vast\rangle.
	\label{eq:bp}
\end{equation}
The matrix elements obtained via this fusion process are explicitly given in terms of $F$-symbols in Ref.~\onlinecite{Levin2005}. 
We remark that this construction can easily be applied to other lattices, by simply splitting each vertex into a connected set of trivalent ones. 

The string-net model with input UFC $\mathcal{C}$ falls into the topological phase with emergent $\mathcal{Z}(\mathcal{C})$ anyons, where $\mathcal{Z}$ denotes the quantum double, or Drinfeld center. 
For the special case where $\mathcal{C}$ is itself an anyon theory (MTC), we have $\mathcal{Z}(\mathcal{C})\cong \mathcal{C} \boxtimes \overline{\mathcal{C}}$, where the bar denotes orientation reversal. 
In this case the string operators can be found easily on the lattice, by simply laying a string from $\mathcal{C}$ over the lattice, for a $\mathcal{C}$ anyon, and another under the lattice, for a $\overline{\mathcal{C}}$ anyon. 
Then the string diagram is resolved using the $R$ moves of the MTC $\mathcal{C}$, and finally reduced to lattice string configurations via $F$ moves. Strings purely over the lattice create the emergent $\mathcal{C}$ anyons, while those purely under the lattice create the $\overline{\mathcal{C}}$ anyons~\cite{koenig2010quantum}. More generally, for any braided UFC, a subset of the string operators can be found by following this procedure. In total generality, when there is no braiding on the UFC $\mathcal{C}$, there is an analogous construction to find all string operators, using a set of ``half-braidings" in place of the $R$ moves. This approach was described in terms of $\Omega$ matrices in Ref.~\onlinecite{Levin2005}.

\section{Notation}
\label{app:notation}

In this work, when considering finite Abelian groups
\begin{align}
    G &\cong \prod_{p\, \text{prime}}\, \prod_{n=1} \Z_{p^{n}}^{d^p_{n}} 
    \, , 
    \\
    \sum_{p,\, n} d_n^p &= k < \infty
    \, ,
\end{align}
we assume a choice of basis has been made for each $\Z_{p^{n}}^{d^p_n}$, with $d^p_n>0$. Concatenating these bases in order of increasing $p$, then $n$ for cyclic groups with the same $p$, defines a basis for $G$ which decomposes it as follows 
\begin{align}
    G 
    &\cong \Z_{p_1^{n_1}} \times \dots \times \Z_{p_k^{n_k}}
    \, .
\end{align}
An arbitrary group element can then be expressed as
\begin{align}
    g = (g_1,\dots,g_k) \, ,
\end{align}
where $g_i=0,\dots,p_i^{n_i}-1$. 
We denote the inverse group element by 
\begin{align}
    \overline{g} = (-g_1,\dots,-g_k) \, .
\end{align}

With the above notation we now define generalized clock and shift matrices. On any finite $G$-graded vector space, with a basis of homogeneous elements $\ket{a_g}$ we define the generalized clock matrix via 
\begin{align}
    \widetilde{Z}^g \ket{a_h} = \prod_{i=1}^{k} \omega_i^{g_i h_i} \ket{a_h}
    \, ,
\end{align}
where $\omega_i$ is a primitive $p_i^{n_i}$-th root of unity. 
For the special case of the vector space $\C[G]$, we have
\begin{align}
    Z^g \ket{h} = \prod_{i=1}^{k} \omega_i^{g_i h_i} \ket{h}
    \, ,
\end{align}
and we can also define a generalized shift matrix via 
\begin{align}
    X^g \ket{h} = \ket{g+h} 
    \, ,
\end{align}
where the group operation is addition of the associated column vectors, modulo $p_i^{n_i}$ for the $i$th entry. 
These operators have the following commutation relations  
\begin{align}
    Z^g X^h = \prod_{i=1}^k \omega_i^{g_i h_i} X^h Z ^g
    \, .
\end{align}

\subsection{Graded string-net constructions}
We consider a family of 2D topological phases that can be represented by string-net models. Physically, this is possible if and only if a topological phase admits a fully gapped boundary, and mathematically they can be described as Drinfeld centers of unitary fusion categories. We further assume that there is a 1-form symmetry group $G$ generated by Abelian bosons ($G$ must be a finite Abelian group). According to Ref.~\onlinecite{SET}, such MTCs can be obtained by gauging a $G$ symmetry of another MTC. For string-net models, the $G$ gauge structure can be realized by an underlying UFC that is $G$-graded~\cite{HeinrichPRB2016,cheng2016exactly}
\begin{equation}
    \mathcal{C}_G=\bigoplus_{g\in G}\mathcal{C}_g \, .
\end{equation}

Without loss of generality we consider the string-net Hamiltonian on a square lattice, with string variables defined on edges, and where each vertex is resolved to be trivalent. The string-net Hamiltonian for $\cat_G$ is given by
\begin{align}
    H_{\cat_G}^{\text{SN}} = - \sum_{v} A_v
    - \sum_{p} \frac{1}{|G|} \sum_{g} B_p^{g}
    \, ,
\end{align}
where $A_v$ projects onto string configurations that satisfy the fusion rules at vertex $v$, and $B_p^{{g}}$ fuses loops from the ${g}$ sector into the boundary of plaquette $p$, see below. Since $G$ is Abelian, we introduce generalized ``clock" operators $\widetilde{Z}^{{g}}_e$ as defined above. 

We remark
\begin{align}
\label{eq:2DZsym}
     \prod_{e \ni v} \widetilde{Z}_e^{\sigma^e_v} A_v = A_v
     \, ,
\end{align}
where $\sigma^e_v=1$ if $e$ points to $v$ and $-1$ otherwise. 
The plaquette terms  form a representation of $G$, as ${B_p^g B_p^h = B_p^{gh}}$, they are given by
\begin{align}
    B_p^g = \sum_{s_g} \frac{d_{s_g}}{\qd_1^2} B_p^{s_g}
    \, ,
\end{align}
where $ B_p^{s_g}$ fuses a loop of $s_g$ string into the edges along the boundary of plaquette $p$. 

Using the notation defined above, we have the commutation relation
\begin{align}
    \widetilde{Z}^g_e B_p^h = \prod_{i=1}^k \omega_i^{ \sigma^p_e g_i h_i} B_p^h \widetilde{Z}_e^g
    \, ,
\end{align}
for $e\in \partial p$, where $\sigma^p_e=1$ if the orientation of $e$ matches that of $p$, and $-1$ otherwise. 
Hence the string operator
\begin{align}
    \widetilde{Z}^g_\gamma : = \prod_{e \cap \gamma} (\widetilde{Z}_e^{g})^{ \sigma^\gamma_e}
    \, ,
\end{align}
is a symmetry when $\gamma$ is a closed loop on the dual lattice, where $\sigma^\gamma_e=1$ if $\gamma$ crosses $e$ in a right handed fashion and
and $-1$ otherwise. These loop operators generate the $G$ 1-form symmetry of the string-net model. 

Furthermore, when $\gamma$ is an open string on the dual lattice, running from plaquette  $\gamma_-$ to  $\gamma_+$, the string operator above creates a $g$ boson at $\gamma_+$ and a $\overline{g}$ boson at $\gamma_-$.

\section{Gauging planar subsystem symmetry}
\label{app:GaugingSubsystems}

In this section we describe the generalized lattice gauging procedure~\cite{WilliamsonPRB2016,VijayPRB2016} for commuting planar subsystem symmetries that have no relations, i.e. no nontrivial products of symmetry operators giving the identity. 

We consider planar subsystem symmetries on the cubic lattice with generators 
\begin{align}
    \prod_{x,y} U_{x,y,z}(g) \, ,
    && 
    \prod_{y,z} V_{x,y,z}(g) \, ,
    &&
    \prod_{x,z} W_{x,y,z}(g) \, ,
\end{align}
forming representations of finite groups $G_{xy},G_{yz},G_{xz},$ on each $xy,yz,xz,$ plane, respectively,  and where the on-site symmetry actions commute 
\begin{align}
[U,V]=[V,W]=[U,W]=0 \, .
\end{align}  
We furthermore focus on the case relevant to our constructions where the subsystem symmetries have no relations, meaning any nontrivial product of symmetry generators cannot give the identity. 
In this case, each plane of the subsystem symmetry can be gauged following the procedure for gauging a global, 0-form symmetry in 2D~\cite{Gaugingpaper}. Since the on-site symmetry actions commute, and there are no relations, the symmetries in orthogonal planes can be gauged sequentially to achieve the same result as simultaneously gauging all at once. Hence it suffices to describe the gauging of one set of planar symmetries and this procedure can then be repeated for the others. 

We proceed to describe gauging the symmetries on $xy$-planes. The first step is to introduce
$\mathbb{C}[G]$ ``gauge" spins on the $\hat{x}$ and $\hat{y}$ edges of the cubic lattice. We label these spins by $e\hat{z}$, for $e\perp \hat{z}$, to distinguish them from gauge spins for the $xz$ and $yz$ subsystem symmetries. 
Projectors implementing the planar Gauss's laws are 
\begin{align}
    P_v^{xy} &= \frac{1}{|G_{xy}|} \sum_{g \in G_{xy}} P_v^{xy}(g) \, ,
    \\
    P_v^{xy}(g) &=  U_v(g) \prod_{e \rightarrow v, e\perp \hat{z}} L_{e\hat{z}}(g)
     \prod_{e \leftarrow v, e\perp \hat{z}} R_{e\hat{z}}(g) \, ,
\end{align}
where $e \rightarrow v~(e\leftarrow v) $ denotes adjacent edges that are oriented towards (away from) $v$. 
Projectors onto zero flux configurations around plaquettes $p\perp \hat{z}$ are given by 
\begin{align}
    F_p^{xy} = \sum_{g_1,g_2,g_3,g_4} & \delta(g_1^{\sigma^p_{e_1}} g_2^{\sigma^p_{e_2}}  g_3^{\sigma^p_{e_3}}  g_4^{\sigma^p_{e_4}}  = 1)
    \nonumber \\ 
    &\pi_{e_1 \hat{z}}(g_1) \pi_{e_2 \hat{z}}(g_2) \pi_{e_3 \hat{z}}(g_3) \pi_{e_4 \hat{z}}(g_4) \, ,
\end{align}
where $\pi_{e \hat{z}}(g) = \ket{g}_{e \hat{z}}\bra{g} $. 
The edges $e_1,e_2,e_3,e_4 \in \partial p$ appear in order starting from an arbitrary vertex in $\partial p$ and following the orientation induced by $p$ along its boundary, with $\sigma^p_{e_i}=1$ if the orientation of $e_i$ matches and $-1$ otherwise. 

To gauge a local Hamiltonian $H=\sum_v h_v$ we first introduce a superoperator that projects operators onto the gauge invariant subspace
\begin{align}
    \mathcal{P}[ \mathcal{O} ] = \sum_{\{ g_v \}} \prod_{v \in S_{\mathcal{O}}} P_v^{xy}(g_v)|_{S_{\mathcal{O}}}~\mathcal{O}~ \prod_{v \in S_{\mathcal{O}}}P_v^{xy}(g_v)|_{S_{\mathcal{O}}}^\dagger \, ,
\end{align}
where $S_{\mathcal{O}}$ is the set of sites in the support of $\mathcal{O}$. 
We use this to define a gauging superopreator for operators on the ``matter" qudits 
\begin{align}
     \mathcal{G}[ \mathcal{O}_m ]  = \mathcal{P}[ \mathcal{O}_m \prod_{e\in T_{\mathcal{O}_m}} \pi_{e \hat{z}}(1) ] \, ,
\end{align}
where $T_{\mathcal{O}_m}$ is a tree, within an $xy$ plane, that contains the vertices in $S_{\mathcal{O}_m}$. 
The gauged Hamiltonian is then 
\begin{align}
    H_{\text{gauged}} = \sum_{v} \mathcal{G}[h_v] - \epsilon \sum_{p} F_p^{xy} - \lambda \sum_{v} P_v^{xy} \, . 
\end{align}
The Gauss's law gauge constraints becomes strict in the limit $\lambda \rightarrow \infty$. 

For the special case that the on-site action of the subsystem symmetry is the regular representation ${U_v(g)=L_{v \hat{z}}(g)\otimes \openone}$, we introduce the following local unitary circuit 
\begin{align}
    LU = \prod_{v} \prod_{e \rightarrow v, e\perp \hat{z}} CL_{v\hat{z},e\hat{z}}
     \prod_{e \leftarrow v, e\perp \hat{z}} CR_{v\hat{z},e\hat{z}} \, ,
\end{align}
where $CL_{1,2}~(CR_{1,2})$ is a left (right) multiplication on qubit 2, controlled by qubit 1. 
Applying this circuit to the gauged Hamiltonian results in a model where the original matter qudits that transform under the regular representation are projected out by the gauge constraints, which become
\begin{align}
    LU P_v^{xy} LU^\dagger = \pi_{v \hat{z}}(1) \otimes \openone \, .
\end{align}

For the relevant case of Abelian groups $G_{xy}$ the Gauss's laws simplify to 
\begin{align}
    P_v^{xy}(g) &=  U_v(g) \prod_{e \rightarrow v, e\perp \hat{z}} X^{g}_{e \hat{z}}
     \prod_{e \leftarrow v, e\perp \hat{z}} X_{e\hat{z}}^{\overline{g}} \, ,
\end{align}
and the flux constraint can be written as 
\begin{align}
    F_p^{xy} = \prod_{e\in \partial p} (Z_{e \hat{z}}^{g})^{\sigma^p_e} \, . 
\end{align}
Throughout the current work we have changed the basis of the gauge spins via an on-site Hadamard transformation resulting in Gauss's law terms
\begin{align}
    H^{\otimes |E|} P_v^{xy}(g)(H^\dagger)^{\otimes |E|} &=  U_v(g) \prod_{e \rightarrow v, e\perp \hat{z}} Z^{g}_{e \hat{z}}
     \prod_{e \leftarrow v, e\perp \hat{z}} Z_{e\hat{z}}^{\overline{g}} \, ,
\end{align}
which are analogous to the $B_c$ terms appearing in our gauged layer Hamiltonians, and 
\begin{align}
    H^{\otimes |E|} F_p^{xy}(H^\dagger)^{\otimes |E|} = \prod_{e\in \partial p} (X_{e \hat{z}}^{g})^{\sigma^p_e} \, , 
\end{align}
which are analogous to the $A_e$ terms appearing in our gauged layer Hamiltonians.

\section{General gauged layers model}
\label{app:GeneralModel}

Here we present the general 3D model constructed from a stack of Abelian $G$-graded string-nets $\double{\cat_G} $ in $xy$ planes along the $\hat{z}$ direction. Such a stack obeys a large symmetry group given by the product of 1-form $G$ symmetries in each layer. This group contains a $G^{L_x}\times G^{L_y}$ subgroup of planar subsystem symmetries generated by elements that are products of the 1-form symmetry in each layer where it is intersected by a dual $xz$ or $yz$ plane. Applying the gauged layers construction leads to the model 
\begin{align}
    H_{\cat_G}^{\text{Frac}} = - \sum_v A_v - \sum_{e\perp \hat{z}} A_e - \sum_{p \perp \hat{z}} B_p' - \sum_c (B_c^{\hat{x}} + B_c^{\hat{y}})
    \, ,
\end{align}
where $A_v$ are string-net fusion-vertex terms, 
\begin{align}
    B_p' = \frac{1}{|G|} \sum_{g} B_p^g  X^g_{p\hat{x}} X^g_{p\hat{y}} \, ,
\end{align}
are gauged string-net plaquette terms, 
\begin{align}
    A_e &= \widetilde{Z}_{e}^\dagger \prod_{p\in \layer_z,\,  e\in\partial p} Z_{p{\hat{e}}}^{\sigma^p_e} \prod_{p \not\in \layer_z,\,  e\in\partial p} Z_{p}^{\sigma^{p}_{e}} \,  ,
\end{align}
are generalized Gauss's law terms, and 
\begin{align}
    B_c^{\hat{x}} &= \prod_{p \in \partial c,\, p \perp \hat{z}} X_{p\hat{x}}^{\sigma^{c}_{p}} \prod_{p \in \partial c,\, p \perp \hat{y}} X_{p}^{\sigma^{c}_{p}} \, ,
    \\
    B_c^{\hat{y}} &= \prod_{p \in \partial c,\, p \perp \hat{z}} X_{p\hat{y}}^{\sigma^{c}_{p}} \prod_{p \in \partial c,\, p \perp \hat{x}} X_{p}^{\sigma^{c}_{p}} \, ,
\end{align}
are zero-flux terms. 
In the above equations we have utilized various notation defined in section~\ref{sec:GaugedLayerConstruction} and appendix~\ref{app:notation}. The symbol $\sigma^c_p$ denotes the relative orientation of a plaquette on the boundary of a cube. We remark that by Poincare duality it is given by $\sigma^c_p=\sigma^{\dl{p}}_{\dl{c}}$ and similarly  $\sigma^p_e=\sigma^{\dl{e}}_{\dl{p}}$, where $\dl{v},\dl{e},\dl{p},\dl{c},$ denotes a dual cube, plaquette, edge, and vertex, respectively.

The ground state consists of a weighted sum of allowed $G$-graded string-net configurations in the 2D $xy$-layers, and Abelian $G$ gauge configurations in the $yz$- and $xz$-layers that may not be closed. With the constraint that an open $g$-labelled string in an $yz$-layer has to end on a string from the $g$ sector in an $xy$-layer, and similar for the gauge fields in the $xz$-layers. 
The unnormalized weights for the string-net configurations are given by the product of the string-net weights in each $xy$-layer.

\subsection{Wilson operators, materialized symmetries and excitations}

There are many generalized Wilson operators that commute with the Hamiltonian, which we proceed to describe. 

First, anyons in the trivial sector of the 2D layers are not affected by the gauging, and hence remain planons.

As we have briefly covered above, anyons in a nontrivial $g\neq 1$ sector become immobile due to the subsystem gauging. In fact, the original string operators no longer commute with the gauged Hamiltonian. The only way to remedy this is to form a loop-membrane operator, i.e. a Wilson loop transporting the anyon from sector $g$ in an $xy$ layer, attached to a membrane of $X^g_{p\hat{x}}X^g_{p\hat{y}}$ in the same layer, or a pair of such loop operators in vertically separated layers, attached by a membrane of $X^g_p$ on the plaquettes in the $yz$ and $xz$ layers between them, the smallest example being
\begin{align}
\label{eq:zcubeterm}
    \prod_{p \in \partial c,\, p \perp \hat{z}} (B_p^g)^{ \sigma^{c}_p} 
    \prod_{p \in \partial c,\, p \not\perp \hat{z}} (X_p^{\overline{g}})^{\sigma^c_p} 
    \,.
\end{align}
Therefore, such anyons can only be created in quadruples, similar to fractons in the X-cube model.

The plaquette operators are not truly membrane operators as we also have Wilson loops given by products of cube terms $B_c^{\hat{x}/\hat{y}}$ over regions in the dual $yz/xz$ planes. 
There are also dual-cage operators on the dual lattice, to see this note $ \prod_{e \ni v,\, e\perp \hat{z}} \widetilde{Z}_e^{\sigma^e_v} $ commutes with the Hamitlonian, see Eq.~\eqref{eq:2DZsym}, and a product of Hamiltonian edge terms on the edges adjacent to a vertex $v$ together with $ \prod_{e \ni v,\, e\perp \hat{z}} \widetilde{Z}_e^{\sigma^e_v} $ leaves a dual-cage term on the dual lattice consisting of $\widetilde{Z}$'s on the plaquettes adjacent to $v$. The smallest such dual-cage is given by 
\begin{align}
    \prod_{e \ni v, \, e \perp \hat{z}} \,
     \prod_{p \ni e, \, p \perp \hat{z}} Z_{p{\hat{e}}}^{\sigma^p_e} \prod_{p \ni e, \, p \not\perp \hat{z}} Z_{p}^{\sigma^{p}_{e}}
    \, .
\end{align}

There are also materialized symmetries~\cite{Kitaev06a,Brown2019} corresponding to particle number parity conservation on certain subsystems due to relations among the Hamiltonian terms, i.e. nontrivial products that give identity. These are generated by: products of $ \prod_{e \ni v,\, e\perp \hat{z}} \widetilde{Z}_e^{\sigma^e_v} $ terms over a layer of constant $z$, products of the dual-cage terms defined above over an $xz$ or $yz$ plane (the relation from an $xy$ plane is dependent on the previous relation), and products of cube terms $B_c^{\hat{x}/\hat{y}}$ over a dual $yz/xz$ plane. These emergent symmetries imply that a vertex excitation which violates $ \prod_{e \ni v,\, e\perp \hat{z}} \widetilde{Z}_e^{\sigma^e_v} $ is a fracton as it is involved in three orthogonal constraints. 
There are further emergent symmetries implied by letting Wilson loops of the type  in Eq.~\eqref{eq:zcubeterm} diverge in size over a dual $xy$ plane. This, together with the other relations derived from the cube terms $B_c^{\hat{x}/\hat{y}}$, implies that excitations of the cube terms are lineons. 
Similarly there is an emergent symmetry implied by letting a Wilson loop of any anyon in the trivial sector diverge in size over an $xy$ plane that implies these excitations are planons.

To summarize, excitations of the vertex terms with $g\neq1$ are fractons. Appropriate pairs of these fractons are planons. 
In particular, excitations of the edge terms are planons that are equivalent to two vertex fractons. 
Excitations of the plaquette terms (and vertex terms with $g=1$) are planons. 
Excitations of the cube terms are lineons. Appropriate pairs of these lineons are planons. We remark that the movement of a fracton in the $xy$ plane is confined as it creates an edge excitation on each edge it traverses.  The $\hat{x}$- and $\hat{y}$- lineons can move in the $\hat{z}$ direction by creating or absorbing an $xy$ planon. Hence the  plaquette planon excitations that are $G$-charges, related to the grading, are equivalent to a pair of lineons. 

\section{Gauged honeycomb layers Hamiltonian}
\label{app:honeycomblayers}

In this section we describe the gauged layer construction for layers of Kitaev's honeycomb model~\cite{Kitaev06a}. Our starting point is the perturbed honeycomb model
\begin{align}
    H-\Delta V = &- \sum_{\langle ij \rangle} J_{\alpha_{ij}} K_{ij} - \sum_{\langle ij \rangle\langle ik \rangle} \Delta K_{ij} K_{ik} 
    \nonumber \\
    & - \sum_{\langle ij \rangle\langle ik \rangle\langle i\ell \rangle} \Delta K_{ij}K_{ik}K_{i \ell} \, ,
\end{align}
see section~\ref{sec:Examples} for our notational conventions. 
We first coarse grain the honeycomb model onto a square lattice, with one $y$-link and two qubits per site. 
The Hamiltonian is a sum of translates of the following terms and their products 
\begin{align}
K_x=
\begin{array}{c}
\xymatrix@!0{%
IX \ar@{-}[r]  & XI
}
\end{array},
&&
K_{y}= YY \, ,
&&
K_{z}=
\begin{array}{c}
\xymatrix@!0{%
IZ  \ar@{-}[d] 
\\
ZI 
}
\end{array},
\end{align}
where, by abuse of notation, $x,y,z$ denotes an $x,y,z$-link. 
The vertical and horizontal $\mathbb{Z}_2$ subsystem symmetries are then generated by 
\begin{align}
    \prod_{i} (ZZ)_{ij}\,, && \prod_{j} (XX)_{ij}\,, 
\end{align}
respectively. 
We next apply a product of Hadamard and controlled-Z gates, $(HH) CZ (HI)$, to every lattice site, bringing the symmetry into a simpler form  
\begin{align}
   \prod_{i} (ZI)_{ij}\,, &&  \prod_{j} (IZ)_{ij}\,.
\end{align}
This takes the terms generating the Hamiltonian to
\begin{align}
K_x'=
\begin{array}{c}
\xymatrix@!0{%
XZ \ar@{-}[r]  & XI
}
\end{array},
&&
K_{y}'=- ZZ \, ,
&&
K_{z}'=
\begin{array}{c}
\xymatrix@!0{%
IX  \ar@{-}[d]  
\\
ZX 
}
\end{array}.
\end{align}

Next, we gauge the $\mathbb{Z}_2$ planar subsystem symmetries formed by products of the linear subsystem symmetries in a stack of the transformed honeycomb model layers. To do so, we first introduce one gauge qubit onto every $\hat{x}$ and $\hat{y}$ edge, and two onto every $\hat{z}$ edge. The gauged Hamiltonian is given by 
\begin{align}
\label{eq:gauged_honeycomb1}
    \widetilde{H} - \Delta \widetilde{V} - \epsilon F - \lambda G \, ,
\end{align}
where the first two terms come from coupling the honeycomb model to the gauge fields, the third term energetically enforces a zero gauge flux constraint, and the fourth term energetically enforces a generalized Gauss's law, which becomes a strict constraint in the limit $\epsilon \rightarrow \infty$. 
The first two terms in Eq.~\eqref{eq:gauged_honeycomb1} are generated by translates of products of the gauged generating terms 
\begin{align}
\widetilde{K}_x=
\begin{array}{c}
\xymatrix@!0{%
XZ \ar@{-}[r] & X  \ar@{-}[r] & XI
}
\end{array},
&&
\widetilde{K}_{y}=- ZZ \, ,
&&
\widetilde{K}_{z}=
\begin{array}{c}
\xymatrix@!0{%
IX  \ar@{-}[d]  
\\
X  \ar@{-}[d]  
\\
ZX 
}
\end{array}.
\end{align}
While the gauge flux constraints are given by translates of 
\begin{align}
\label{eq:honeycomblayersflux}
\begin{array}{c}
\xymatrix@!0{%
&& 
\\
& IX \ar@{-}[dl] \ar@{-}[ur]  & X \ar@{-}[u] \ar@{-}[d]
\\
 &&
\\
X  \ar@{-}[u]\ar@{-}[d]  & IX \ar@{-}[dl] \ar@{-}[ur]
\\
&&
}
\end{array},
&&
\begin{array}{c}
\xymatrix@!0{%
&& & X  \ar@{-}[l]\ar@{-}[r] & 
\\
& XI \ar@{-}[dl] \ar@{-}[ur]  & & XI \ar@{-}[ur] \ar@{-}[dl] 
\\
  & X  \ar@{-}[l]\ar@{-}[r] & 
}
\end{array},
\end{align}
and the generalized Gauss's laws are 
\begin{align}
\begin{array}{c}
\xymatrix@!0{%
& Z & IZ 
\\
  & IZ \ar@{-}[dl] \ar@{-}[ur]  \ar@{-}[u]\ar@{-}[d] & 
\\
IZ & Z &
}
\end{array},
&&
\begin{array}{c}
\xymatrix@!0{%
 && ZI 
\\
Z  & ZI \ar@{-}[dl] \ar@{-}[ur]  \ar@{-}[l]\ar@{-}[r] & Z
\\
ZI & &
}
\end{array}.
\end{align}

The gauged Hamiltonian is equivalent to a model on the edge qubits alone via a local unitary circuit 
\begin{align}
U= \prod_{v} \prod_{e \ni v, e \parallel \hat{x}} C X_{e,v_1} \prod_{e \ni v, e \parallel \hat{y}}  C X_{e,v_2} \prod_{e \ni v, e \parallel \hat{z}} C X_{e_1,v_1} C X_{e_2,v_2} ,
\end{align}
where $e_{1,2},v_{1,2},$ denote the left and right qubits on vertices and $\hat{z}$ edges, respectively. 
The final Hamiltonian is given by 
\begin{align}
    \widetilde{H}' - \Delta \widetilde{V}' - \epsilon F \, ,
\end{align}
where the first two terms are generated by translates of products of 
\begin{align}
\widetilde{K}_x'=
\begin{array}{c}
\xymatrix@!0{%
& Z & IZ 
\\
  &  \ar@{-}[dl] \ar@{-}[ur]  \ar@{-}[u]\ar@{-}[d] \ar@{-}[r]\ar@{-}[l] & X
\\
IZ & Z &
}
\end{array},
&&
\widetilde{K}_{y}'=-
\begin{array}{c}
\xymatrix@!0{%
& Z & ZZ 
\\
  Z  &  \ar@{-}[dl] \ar@{-}[ur] \ar@{-}[u]\ar@{-}[d] \ar@{-}[l]\ar@{-}[r] & Z
\\
ZZ & Z &
}
\end{array} ,
\nonumber \\
\widetilde{K}_{z}'=
\begin{array}{c}
\xymatrix@!0{%
 & X & ZI 
\\
Z  &  \ar@{-}[dl] \ar@{-}[ur] \ar@{-}[u]\ar@{-}[d] \ar@{-}[l]\ar@{-}[r] & Z
\\
ZI & &
}
\end{array},
\end{align}
and the third term is the same as in Eq.~\eqref{eq:gauged_honeycomb1}. The vertex qubits can be ignored as they are projected onto $\ket{00}$ by the Gauss's law operators after the local unitary transformation maps them to $IZ,ZI$.

\end{document}